\documentclass[smallextended]{svjour3}       
\smartqed  
\usepackage{amsfonts,amsmath,amssymb}
\usepackage{graphicx}
\usepackage{cite}
\usepackage{xcolor}
\usepackage{soul}
\usepackage[left]{lineno}
\definecolor{green}{rgb}{0,0.5,0}

\begin{document}

\title{Fragmentation in Frustrated Magnets}
\subtitle{A review}


\author{Elsa Lhotel \and Ludovic D.C. Jaubert \and Peter C.W. Holdsworth}

\authorrunning{Lhotel, Jaubert, Holdsworth} 

\institute{E. Lhotel \at
              Institut N\'eel, CNRS - Universit\'e Grenoble Alpes, 38042 Grenoble, France.\\
              \email{elsa.lhotel@neel.cnrs.fr}\\
              \and
              L.D.C. Jaubert \at
              CNRS, Universit\'e de Bordeaux, LOMA, UMR 5798, 33405 Talence, France.\\
              \email{ludovic.jaubert@u-bordeaux.fr}\\
              \and
              P.C.W. Holdsworth \at
              Universit\'e de Lyon, ENS de Lyon, Universit\'e Claude Bernard, CNRS, Laboratoire de Physique, 69342 Lyon, France.
              \email{peter.holdsworth@ens-lyon.fr}\\
              }


\maketitle

\begin{abstract}
Spin liquids are exotic phases of matter that often support emergent gauge fields and quasi-particle excitations. While spin liquids are commonly known for remaining disordered, their definition has been extended to include phases with broken symmetry corresponding to (partial) long-range order, such as chiral and nematic spin liquids for example. For Coulomb spin liquids, this ordering can be quantitatively understood via a Helmholtz decomposition between divergence-free and divergence-full terms. This phenomenon has been coined fragmentation, where spin degrees of freedom fragment into two components; the fluctuating disordered part and the ordered one. In this review, we will cover the theoretical and experimental aspects of this growing field, in particular its relation to magnetic monopoles in spin ice, its phase diagram and the possibility to observe it in solid-state crystal and artificial networks.\\
\end{abstract}

\section{Introduction}

Spin liquids are unconventional phases of matter where frustration prevents magnetic order down to the lowest temperatures \cite{Savary16b,Knolle18a}, \textcolor{black}{opening a window for new collective phenomena}. {\color{black} But while the absence of order has the merit of summarising their most noticeable property, and to be reasonably straightforward to identify experimentally, it is not really satisfactory to define something by what it is not. The caveat of a definition by negation is that it tends to group together concepts which can be fairly different, linked here by the thread of magnetic disorder. To improve the description, one needs to characterise disorder, which is a counter-intuitive task in a field -- condensed matter -- dominated by Landau theory, since disorder bears no broken symmetries.} This is where the emergence of quasi-particles and gauge field theory becomes useful.

Emergence here means renormalisation from discrete to higher symmetry, and leads to an elegant simplification of the initial problem \cite{Anderson72a,NPhys16a}, providing a theoretical framework for the understanding of complex many body physics \cite{Wen04a}. For example magnetic excitations of spin liquids often take the form of quasi-particles, such as 
Majorana fermions in Kitaev materials \cite{Kitaev06a} and magnetic monopoles in spin ice \cite{Castelnovo08,Ryzhkin05}. In particular, the latter description has offered an unlikely cadre for the study of Coulomb physics and emergent electromagnetism \cite{huse03a,isakov04b,henley05a,Henley10,rehn16a}.
 
\begin{figure}[t]
\centering\includegraphics[width=12cm]{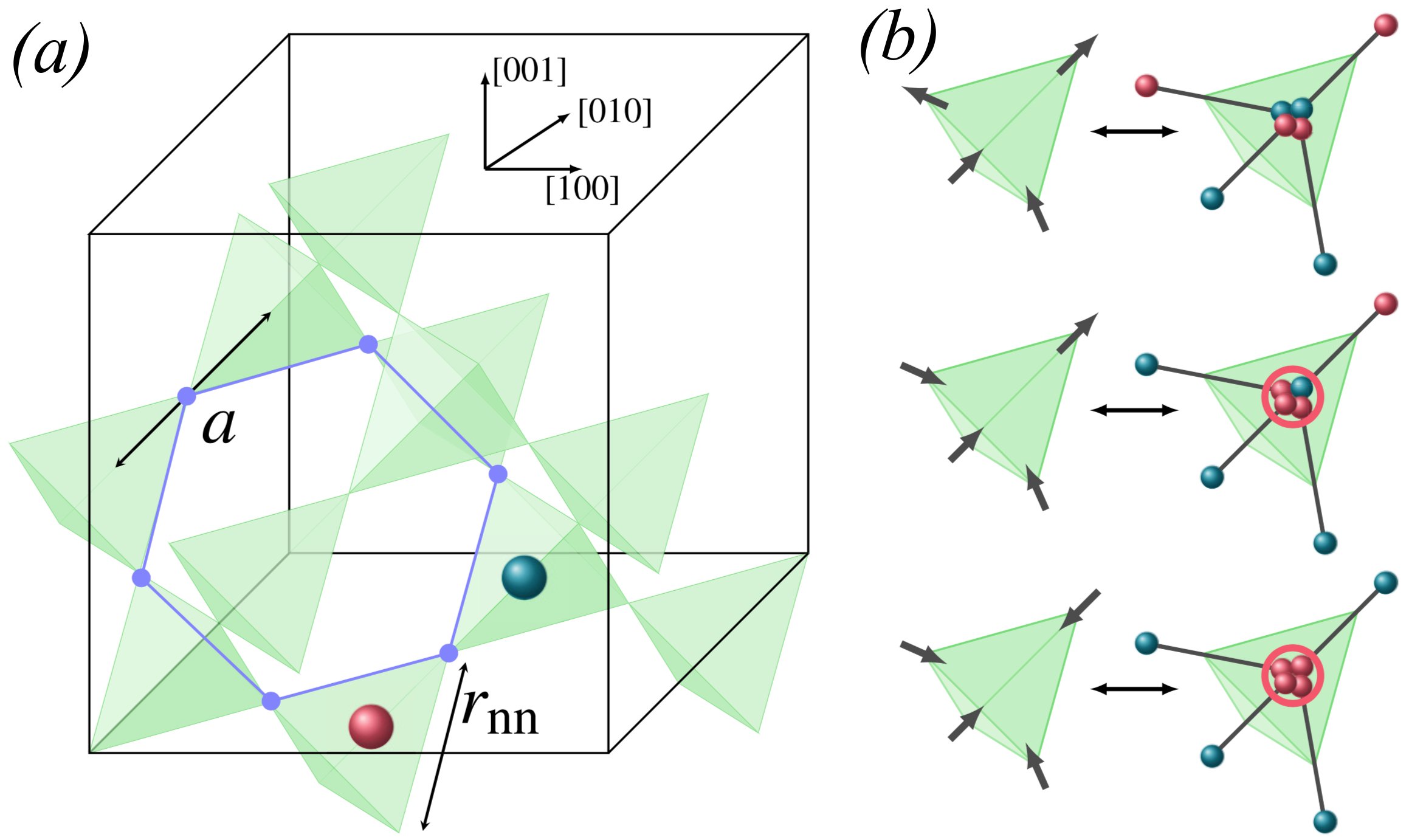}
\caption[Emergence of monopoles]{
\textbf{$(a)$ Pyrochlore lattice:}  cubic unit cell of a pyrochlore lattice. The centre of the tetrahedra form a diamond lattice whose bipartite nature particularises up and down tetrahedra, with a red and blue sphere respectively. The diamond lattice forms the backbone for magnetic monopoles.
\textbf{$(b)$ Dumbbell model:} The magnetic dipoles are replaced by a needle carrying a dumbbell of charge - small red and blue spheres. 2 in - 2 out tetrahedra (top) are charge neutral, 3 in - 1 out and 3 out - 1 in carry a single-charge monopole (centre), 4in and 4out carry a double monopole (bottom). Note that a monopole is a composition of several dumbbell charges allowing monopole fractionalisation. Reproduced from Ref.~\cite{Raban19}.
}
\label{fig1}
\end{figure}

Spin ice is a canonical frustrated model made of Ising spins on the three-dimensional (3D) pyrochlore lattice [Fig.~\ref{fig1}] \cite{Harris97,bramwell01b}. The low temperature state of spin ice, with associated extensive Pauling entropy \cite{Harris97,Ramirez99}, constitutes an effective vacuum of divergence free fluctuations from which quasi-particles carrying magnetic charge are thermally excited. This can be seen as a Helmholtz decomposition of the emergent magnetostatic field, such that the moments appear to fragment into two magnetic fluids \cite{Brooks14}, the first providing deconfined Coulomb particles, the second giving a perpetually fluctuating background with topological properties. This picture shares many features with the electrostatic description of Kosterlitz-Thouless spin systems \cite{Kosterlitz73a,Villain75a} in which vortices and spin waves provide an equivalent set of electrostatic fluids \cite{Faulkner15a}.

The goal of this review is to show how magnetic fragmentation provides a new platform going beyond the traditional picture of spin liquids. Indeed, while the fluctuating background is essentially a Coulomb spin liquid, the magnetic charges can be manipulated independently and even crystallise. Fragmentation thus allows for the stabilisation of a new phase in which {\color{black} magnetic} order coexists with a fluctuating spin liquid with ferromagnetic correlations \cite{Brooks14,Borzi13}, which has recently been observed in several experimental contexts: Nd$_{2}$Zr$_{2}$O$_{7}$ \cite{Petit16b}, Ho$_{2}$Ir$_{2}$O$_{7}$ \cite{Lefrancois17}, Dy$_3$Mg$_2$Sb$_3$O$_{14}$ \cite{Paddison16}, artificial lattices \cite{Canals16}...\\

\section{Theory of magnetic fragmentation}
\label{sec:theory}
\subsection{Introduction to spin ice}

The localized magnetic moments of spin ice systems~\cite{Harris97} form a corner sharing network of tetrahedra on a pyrochlore lattice, as shown in Fig.~\ref{fig1}.$(a)$. The physics of spin ice is commonly described by the dipolar spin ice Hamiltonian (DSI) \cite{denHertog00,Melko04} made up of both exchange and dipolar terms
\begin{eqnarray}
\mathcal{H}\;=\;-J\,\sum_{\left<i,j\right>}\mathbf{S}_{i}\cdot\mathbf{S}_{j}\;+\;
D\, r_{nn}^{3}\, \sum_{i < j}\left[
\frac{\mathbf{S}_{i}\cdot\mathbf{S}_{j}}{\left|\mathbf{r}_{ij}\right|^{3}}-
\frac{3\left(\mathbf{S}_{i}\cdot\mathbf{r}_{ij}\right)\left(\mathbf{S}_{j}\cdot\mathbf{r}_{ij}\right)}{\left|\mathbf{r}_{ij}\right|^{5}}\right]
\label{eq1}
\end{eqnarray}
where $r_{nn}$ is the distance between nearest neighbour spins and $\vec r_{ij}$ is the vector between two pyrochlore sites $i,j$. The first sum runs over nearest-neighbour pairs labeled $\left< ... \right>$ while the second sum runs over all pairs of spins. $\mathbf{S}_{i}=\sigma_{i}\vec e_{i}$ is an Ising like spin of unit length, $\sigma_{i}=\pm 1$, constrained to point along the easy axis $\vec e_{i}$ joining the centres of the two adjacent tetrahedra. There are four sites in a unit cell, which we take here to be an ``up'' tetrahedron as shown in Fig.~\ref{fig1}.$(b)$ and thus four different easy axes
\begin{eqnarray}
\vec e_{i}\in \left\{
\frac{1}{\sqrt{3}}\begin{pmatrix}+1\\+1\\+1\end{pmatrix},
\frac{1}{\sqrt{3}}\begin{pmatrix}-1\\-1\\+1\end{pmatrix},
\frac{1}{\sqrt{3}}\begin{pmatrix}-1\\+1\\-1\end{pmatrix},
\frac{1}{\sqrt{3}}\begin{pmatrix}+1\\-1\\-1\end{pmatrix}
\right\}
\label{eq:easyaxis}
\end{eqnarray}
Typically for spin ice materials Dy$_{2}$Ti$_{2}$O$_{7}$ and Ho$_{2}$Ti$_{2}$O$_{7}$ the magnetic moment is $m\approx 10\mu_{B}$ and the coupling constants are on the $1$ K energy scale; for Dy$_{2}$Ti$_{2}$O$_{7}$ $J \approx3.72$ K and $D = \dfrac{\mu_{0} m^{2}}{4\pi r_{nn}^3} \approx 1.41$ K ~\cite{denHertog00}. 

The properties of spin ice can be described to a first approximation by cutting off  the dipole interaction at nearest neighbour, leading to the nearest neighbour spin ice model (NNSI)
\begin{equation} 
\mathcal{H_{NN}}= \;-3J_{\rm eff}\,\sum_{\left<i,j\right>}\mathbf{S}_{i}\cdot\mathbf{S}_{j}=\;J_{\rm eff}\,\sum_{\left<i,j\right>}\sigma_{i}\sigma_{j},
\end{equation}
with $J_{\rm eff}=\frac{5D}{J}+\frac{J}{3}$. The spin ice regime $J_{\rm eff}>0$ gives a frustrated ferromagnet in which the lowest energy state of a single tetrahedron is six fold degenerate, with two spins pointing inwards and two outwards, as shown in Fig.~\ref{fig1}.$(b)$. Tiling these configurations together in the pyrochlore lattice gives an extensive band of degenerate ground states - the Pauling states \cite{Pauling35a}- sharing the same phase space as the protons in the cubic phase of ice, giving spin ice its name \cite{Harris97}. As a consequence, there is no phase transition in this model. As temperature goes to zero, the system progressively enters a collective paramagnetic state which violates the third law of thermodynamics in retaining the Pauling entropy of the ice rules states.

The local constraints can be written as a divergence free condition of the coarse-grained magnetisation, or magnetic moment density, $\vec M$ satisfying Maxwell-Gauss law $\vec \nabla \cdot\vec M=0$ \cite{isakov04b,henley05a}. In other words, $\vec M$ behaves as the curl of an emergent gauge field $\vec A$, $\vec M \equiv \vec\nabla \wedge \vec A$. At this level of approximation, with an effective Hamiltonian, which is pure exchange, $\vec M$ can be considered as a field built from the dimensionless spins. These develop emergent dipolar correlations because of the constraints \cite{isakov04b} in what is referred to as a Coulomb phase \cite{Henley10}. A consequence of these correlations are characteristic pinch points in simulated neutron scattering patterns \cite{youngblood81a,Fennell09}. 

When going beyond the nearest-neighbour cutoff, the pyrochlore lattice possesses a remarkable symmetry property which ensures that the long range part of the dipolar interaction is almost perfectly screened within the ensemble of ice-rule states \cite{denHertog00,isakov05a}. This screening means that the DSI inherits these states in the form of a quasi-degenerate manifold of lowest energy states respecting the divergence-free condition. The degeneracy is ultimately lifted through terms of quadrupolar and higher order giving an ordered ground state at a temperature in the 0.1 K range \cite{Melko04}. However, above this temperature, the extensively degenerate manifold is essentially recovered so that the physics of real spin ice materials is largely dictated by the emergent gauge field of the NNSI. In particular, the pinch point scattering patterns characteristic of the dipolar correlations of the lattice gauge field are spectacularly reproduced in experiments \cite{youngblood81a,bramwell01b,Fennell09}. 

However, there is a subtle difference between the two models. With the inclusion of real dipolar interactions comes the inclusion of real magnetic flux and a dimension full magnetisation $\vec M$ which, thanks to the above symmetry properties lies almost perfectly on top of the emergent field. As a consequence, the topological defects which appear in the emergent field of the NNSI \cite{Ryzhkin05} are dressed, in the DSI with real magnetic flux and carry real magnetic charge. These excitations are magnetic monopoles \cite{Castelnovo08} and the quasi-degenerate band of Pauling states plays the role of a vacuum from which they are excited (see  Fig.~\ref{fig1}.$(b)$). Hence, as shown in detail below, while the topological defects of the NNSI have entropic interactions, \textcolor{black}{they behave like magnetic monopoles with real Coulomb interactions in the DSI.}


{\color{black} Magnetic monopoles emerge naturally as an approximation to the DSI Hamiltonian via the so-called the dumbbell model}\cite{Castelnovo08}: extending the point dipoles to infinitesimally thin magnetic needles lying along the easy axes linking the centres of adjoining tetrahedra (see Fig.~\ref{fig1}) which constitute a diamond lattice. The needles carry dumbbells of charges at each end which touch at the diamond lattice sites. For each dumbbell, there is a positive and a negative charge, reproducing the magnetic moment of the original point dipole. By construction, the ground-state ensemble of ice-rule states with two spins pointing in and two spins pointing out is thus charge neutral for all tetrahedra, recovering the screening of the long-range dipolar interactions mentioned above. The quadrupolar and higher order perturbations of the screening correspond to the approximation made when mapping point dipoles to extended dumbbells. Reversing the orientation of a needle (or spin) breaks the ice rules on a pair of neighbouring diamond sites, creating two local magnetic charges: a positive and a negative one. \textcolor{black}{The dumbbell model is the lattice gauge theory of the DSI. Describing the physics of spin ice as a gas of magnetic charges offers an elegant intuition of its thermodynamics (heat capacity, field-induced phase transitions ...) and dynamics. In this review, this picture will come especially handy since the long-range magnetic order can often be described as a monopole crystallisation.}

The magnetic moment $m\;\vec S_{j}$ on the site $j$ of the pyrochlore lattice forms an element of a lattice field $M_{IJ}$ on the bonds of the diamond diamond lattice of tetrahedron centres, connecting the two adjacent diamond sites, $I$ and $J$. The lattice field $M_{IJ}$ is the magnetisation flux channelled through the needle. Formally $M_{IJ}=\vec M \cdot d\tilde{\vec S}$, where $\vec M$ (in the dumbbell model) is defined uniquely within the needle and $d\tilde{\vec S}$ is the infinitesimal needle cross section pointing out of the tetrahedron. The $M_{IJ}$ thus have units of magnetic charge and take the form
\begin{eqnarray}
M_{IJ}=(\vec S_{j}\cdot\vec e_{j})\frac{m}{a}\eta_I=\sigma_{j}\frac{m}{a}\eta_I,
\label{eqdefMIJ}
\end{eqnarray}
where $a$ is the nearest neighbour distance on the diamond lattice (see Fig.~\ref{fig1}.$(a)$) and where $\eta_I=1(-1)$ for an up(down) tetrahedron, ensuring that $M_{IJ}=-M_{JI}$. \textcolor{black}{An artistic view of the magnetisation flux $M_{IJ}$ is given in Fig.~\ref{fig:frageq}.} The integral form of Gauss' law then takes the discrete form,
\begin{equation}
\sum_{J=1}^{4} M_{IJ}=-Q_I,
\label{eqdisGauss}
\end{equation}
where the sum runs over the four nearest neighbours $J$ on the diamond lattice. $Q_I\in\{0,\pm Q, \pm 2Q\}$ is the magnetic charge, with  $Q=2m/a$, the monopole charge. Note that with this sign convention (complicated as it may be!) in which an inward pointing field element carries a minus sign, the $M_{IJ}$ satisfy the requirements of both the emergent magnetostatics of the gauge field and the conventional magnetostatics of the real magnetic problem, as shown in detail below.


Within the dumbbell approximation, the Hamiltonian (\ref{eq1}) becomes (up to a constant)
\begin{equation}
\mathcal{H}={1\over{2}}\sum_{I\ne J}{\mu_0 Q_I Q_J\over{4\pi\; r_{IJ}}} -\mu N -\mu_{2} N_2,
\label{eq2}
\end{equation} 
where the sum runs over all pairs of diamond sites $I,J$. $\mu<0$ and $\mu_2=4\mu$ are chemical potentials for the creation/annihilation of single and double charged monopoles (see Fig.~\ref{fig1}.$(b)$). Whithin this mapping, the problem is reformulated as a lattice Coulomb fluid in the grand ensemble and $\mu$ and $\mu_{2}$ can be calculated for each material from the parameters $D$ and $J$ of the corresponding DSI model \cite{Castelnovo08}

\begin{eqnarray}
|\mu|=\frac{2J}{3}+\frac{8}{3}\left[1+\sqrt{\frac{2}{3}}\right]D=\frac{|\mu_{2}|}{4}
\label{eq:mu}
\end{eqnarray}
%

\subsection{Helmholtz decomposition}
\label{sec:Helmholtz}
\begin{figure} [t]
\centering\includegraphics[width=12cm]{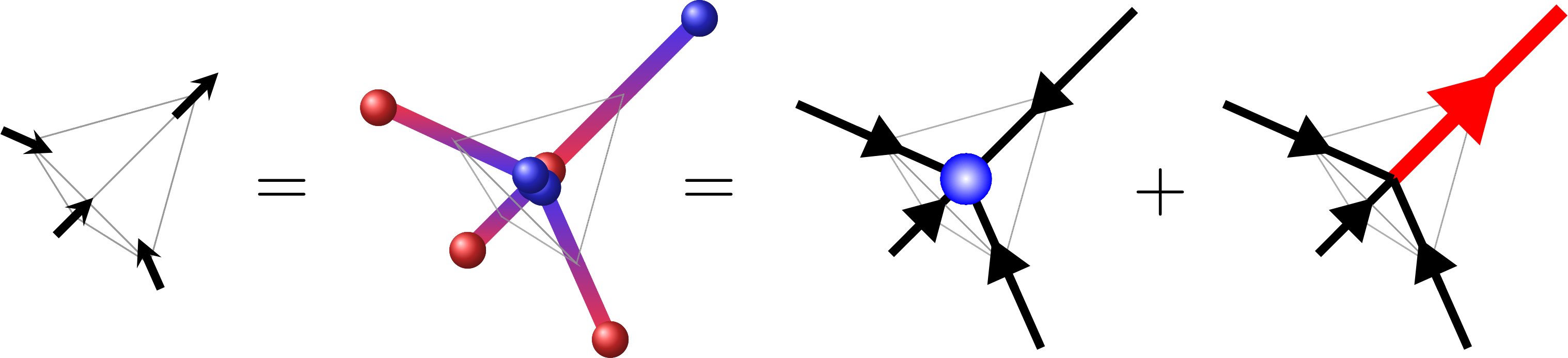}
\caption[Helmholtz decomposition]{\textcolor{black}{\textbf{Helmholtz decomposition:} Spin configuration for a 3 in - 1 out tetrahedron (left) with its dumbbell representation made of $\pm Q/2=\pm \frac{m}{a}$ magnetic charges (middle) and their Helmholtz decomposition into divergence-full and divergence-free elements (right). The divergence-full term contains a positive magnetic charge. Adapted from \cite{Brooks14}}
\label{fig:frageq}}
\end{figure}

In a magnetostatic problem, the coarse grained magnetisation satisfies Gauss' law in the form $\vec\nabla \cdot \vec M=-\rho_{\mathrm{m}}$ where $\rho_m$ is the concentration of induced magnetic charge. Given the emergent properties  of spin ice, in the absence of boundaries and disorder, finite $\rho_m$ is due to the presence of magnetic monopoles and the study of the real magnetostatics gives access to this emergence. 
Is it then possible to quantify by how much the divergence-free condition is broken by the magnetic monopoles ? The answer comes via a Helmholtz decomposition:
\begin{equation}
\vec M = \vec M_{\mathrm{m}}+ \vec M_{\mathrm{d}}= \vec\nabla \psi(r) + \vec\nabla\wedge\vec A.
\end{equation}
The magnetisation $\vec M$ is \textit{fragmented} \cite{Brooks14} into two contributions. The first, $\vec M_{\mathrm{m}}$, falls on the gradient of a scalar potential and provides the magnetic charge. The second, $\vec M_{\mathrm{d}}$, a dipolar field, can be represented as the curl of a vector potential; it is divergence-free and is at the origin of the Coulomb phase for states obeying the ice rules, for which $\vec M_{\mathrm{m}}=0$. In this language, breaking the ice rules leads to the conversion of $\vec M$ from the divergence-free field $\vec M_{\mathrm{d}}$ to the divergence-full field $\vec M_{\mathrm{m}}$. This conversion is complete at the microscopic level, for the double monopoles (see Fig.~\ref{fig1}.$(b)$), but is only partial in single charged monopoles, so that one has the coexistence of two complementary fields or fluids, ($\vec M_{\mathrm{d}}$, $\vec M_{\mathrm{m}}$) \textcolor{black}{[see Fig.~\ref{fig:frageq} for a schematic representation of the Helmholtz decomposition of an isolated positive magnetic charge]}.

It is important to stress at this point that, in spin ice the emergence and subsequent decomposition of $\vec M$ into two orthogonal fluids is not just a coarse grained phenomenon. It occurs on the microscopic scale and the dumbbell model provides a mathematical framework for this.
Each element, $M_{IJ}$, fragments into a monopolar and dipolar part, $M_{IJ}=M_{IJ}^{\rm m}+M_{IJ}^{\rm d}$ such that $\sum_J M_{IJ} ^{\rm m}=-Q_I$. As the magnetic moments are of fixed size, there is a microscopic constraint on each element
\begin{eqnarray}
(M_{IJ}^{\rm m}+M_{IJ}^{\rm d})\frac{a}{m}=\pm1\label{constraint-1}, 
\end{eqnarray}
in addition to the global orthogonality condition
\begin{equation}
\sum_{I>J} (M_{IJ}^{\rm m})(M_{IJ}^{\rm d})=0\label{constraint-2}.
\end{equation}
Transformed into vectors, $\vec M_{IJ}=(\vec e_i\eta_I) M_{IJ}$ and then into reciprocal space, with wave vector $\vec q$ restricted to the first Brillouin zone, $\vec M_{\mathrm{m}}(\vec q)$ and $\vec M_{\mathrm{d}}(\vec q)$ are respectively parallel and perpendicular to $\vec q$ giving ``longitudinal'' and ``transverse'' components to $\vec M(\vec q)$.

The equations of constitutive magnetostatics follow: treating the  needles as lossless conduits of magnetic flux one can define lattice elements of magnetic intensity
\begin{eqnarray}
 H_{IJ}=-M_{IJ}^{\rm m},
\end{eqnarray}
and magnetic induction
\begin{eqnarray}
B_{IJ}=\mu_0 (M_{IJ}+H_{IJ})= \mu_0 M_{IJ}^{\rm d},
\end{eqnarray}
giving Maxwell's equation $\sum_J B_{IJ} =0$ and, following (\ref{constraint-1}) a local field along each element, 
$B_{IJ}-\mu_0M^{\rm m}_{JI}$ of constant amplitude. The elusive Dirac strings that maintain Maxwell's equation \cite{Castelnovo08} are thus incorporated into the network of field elements. 

\begin{figure} [t]
\centering\includegraphics[width=12cm]{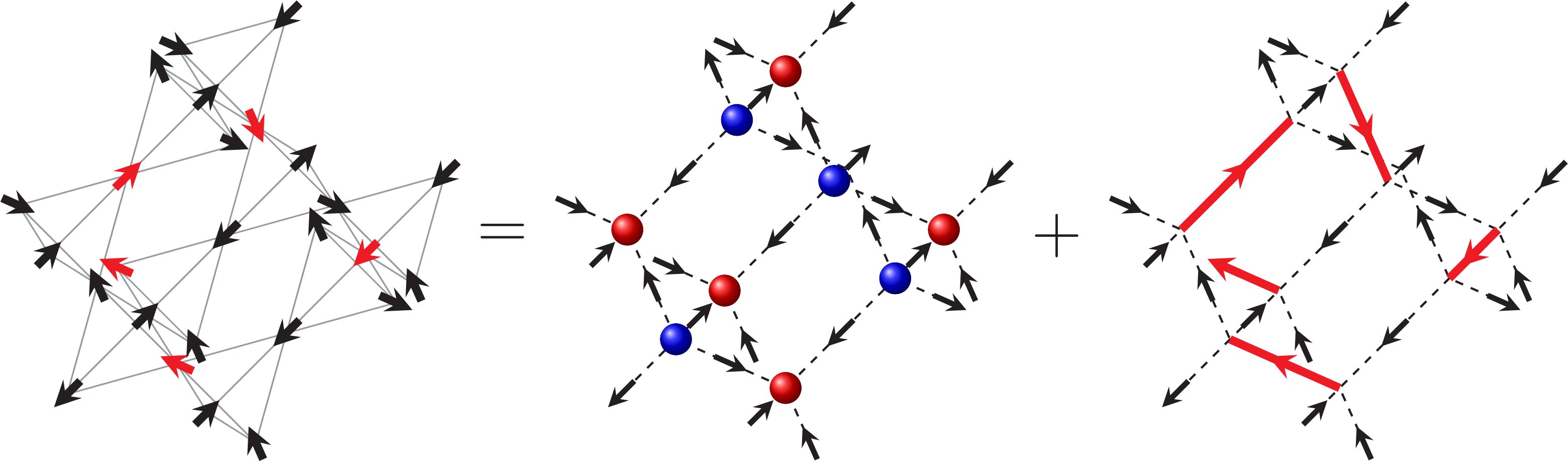}
\caption[Fragmented spin liquid]{\textbf{Fragmented spin liquid on the pyrochlore lattice:} (left) Spin configuration with alternatively 3 in - 1 out and 3 out - 1 in tetrahedra. The red spins are the minority spins. Following Eq.~(\ref{eq:frag}), this spin configuration can be (Helmholtz) decomposed into a monopole crystal (centre) and a hard-core dimer model (right) on the diamond lattice \cite{Brooks14}. \textcolor{black}{The divergence-free field emerging from the dimer model, despite being a Coulomb phase, is different from the traditional one in the ground state of spin ice with only 2 in- 2 out tetrahedra.} The red arrows on the right panel carry three times more flux than the black ones in the middle and right panels.
\label{fig:frag}}
\end{figure}

Concretely, the magnetisation flux of Eq.~(\ref{eqdefMIJ}) can be written in vector form around a given diamond site $I$; for example for an isolated monopole of positive charge (a north pole) with three inward and one outward spin
\begin{eqnarray}
[M_{I}]=(-1,-1,-1,1)\;\frac{Q}{2}.
\label{separation}
\end{eqnarray}
Its fragmentation gives \cite{Brooks14} \textcolor{black}{[Figs.~\ref{fig:frageq} and \ref{fig:frag}]}
\begin{eqnarray}
[M_{I}]=\left(-{1\over{2}},-{1\over{2}},-{1\over{2}},-{1\over{2}}\right)\;\frac{Q}{2} +
\left(-{1\over{2}},-{1\over{2}},-{1\over{2}},{3\over{2}}\right)\;\frac{Q}{2}.
\label{eq:frag}
\end{eqnarray}
The first term satisfies Gauss' law for the magnetic charge; the second term satisfies the discrete version of a divergence-free field. Notice that the magnitudes of $M_{IJ}^{\rm m}$ and $M_{IJ}^{\rm d}$ can exceed $m/a$ as long as the constraint (\ref{constraint-1}) is satisfied; for example here one field element has $M_{IJ}^{\rm m}=-m/2a$ and $M_{IJ}^{\rm d}=3m/2a$.

As the Coulomb interaction is long ranged, moving away from the monopole one can find non-zero $M_{IJ}^{\rm m}$ in a volume containing no charge. These elements can be calculated by solving Poisson's equation from which $M_{IJ}^{\rm d}$ can be calculated from the constraint (\ref{constraint-1}). 
Such a decomposition into $[M_{IJ}^{\rm m}]$, $[M_{IJ}^{\rm d}]$ is possible for any spin ice configuration, regardless of the concentration of monopoles and double monopoles{\color{black}. Nevertheless,} 
for periodic boundaries it is actually multi-valued through topological sector fluctuations \cite{Jaubert13a}, which are allowed even in the presence of charges, if divergence free pathways span the system. This procedure has been verified in the analogous Kosterlitz-Thouless environment \cite{Faulkner15a} and an approximate algorithm for the decomposition has recently been developed for spin ice \cite{Slobinsky19a}.

\begin{figure}[t]
\centering\includegraphics[width=12cm]{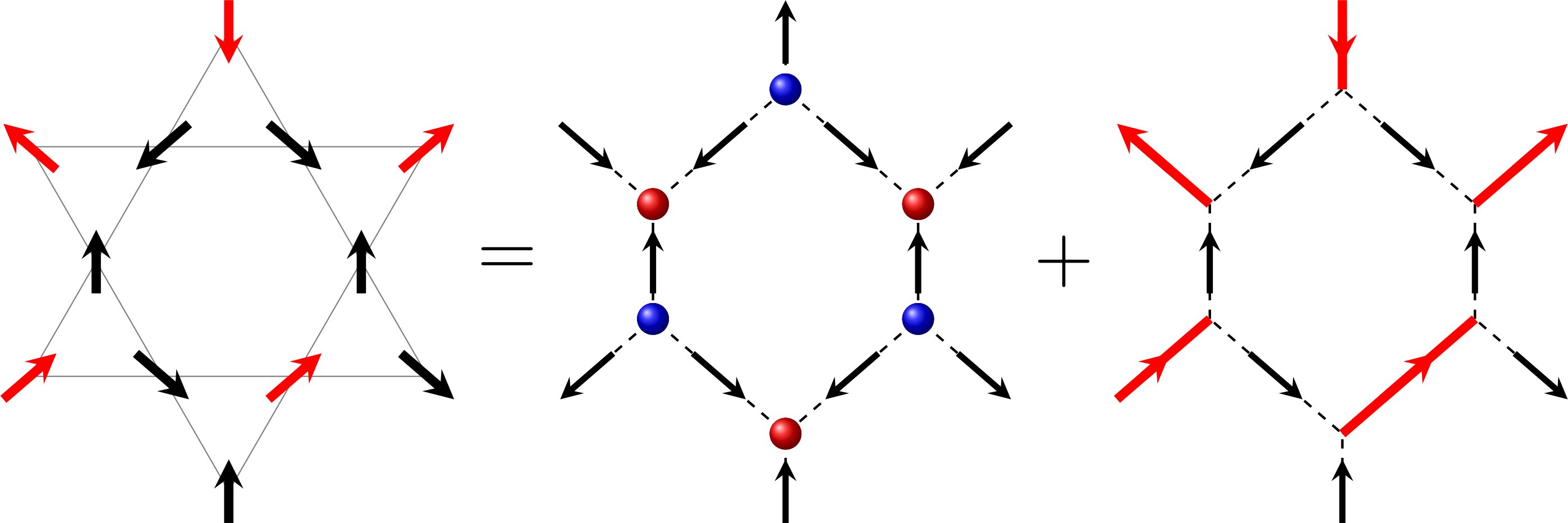}
\caption[Fragmented spin liquid]{\textbf{Fragmented spin liquid on kagome ice:} (left) Spin configuration with alternatively 2 in - 1 out and 2 out - 1 in triangles. The red spins are the minority spins. Following Eq.~(\ref{eq:fragk}), this spin configuration can be (Helmholtz) decomposed into a monopole crystal (centre) and a hard-core dimer model (right) on the honeycomb lattice \cite{Brooks14}. The red arrows on the right panel carry two times more flux than the black ones in the middle and right panels.
\label{fig:fragk}}
\end{figure}

\textcolor{black}{The notion of a fragmented crystal can be extended to two dimensions, in particular on the kagome lattice, when spins are constrained to point in or out of the triangles [Fig.~\ref{fig:fragk}]. In the presence of an effective ferromagnetic nearest neighbour interaction, the ground state is extensively degenerate with six possible states per triangle; two spins pointing in and one out, or the inverse\footnote{This phase is analogue to the standard kagome Ising antiferromagnet with collinear Ising spins.}. Adding dipolar interactions stabilises a novel phase at lower temperatures with broken $\mathbb{Z}_{2}$ symmetry \cite{Moller09,Chern11,Canals16}. The broken symmetry corresponds to the fact that only three states are now allowed per triangle; 2 in - 1 out for down triangles and 2 out - 1 in for up ones, or vice-versa [Fig.~\ref{fig:fragk}]. It is the so-called kagome ice phase (KI). This phase is also sometimes labeled ``KI 2'' as opposed to ``KI 1'' for the above mentioned ground state without the broken symmetry.} The triangle incorporates a magnetic charge equal to half the monopole charge on pyrochlore ($+Q/2$) and the ice rule can be fragmented to give
\begin{eqnarray}
[M_{I}]=(-1,-1,1)\;\frac{Q}{2}=\left(-{1\over{3}},-{1\over{3}},-{1\over{3}}\right)\;\frac{Q}{2} +
\left(-{2\over{3}},-{2\over{3}},+{4\over{3}}\right)\;\frac{Q}{2}.
\label{eq:fragk}
\end{eqnarray}
The ground state forms a charge crystal discussed in more detail below. Interestingly, taking the unit of charge to be that of the monopole, the crystal provides an example of geometrically driven charge fractionalisation \cite{Fulde02a}. Anywhere in the phase diagram, the monopole configuration can be extracted by subtracting the charge crystal from the total magnetic charge configuration. The contribution of an isolated monopole to $[M_{I}^{\rm m}]$ is then $\left(-{2\over{3}},-{2\over{3}},-{2\over{3}}\right)\;\frac{Q}{2}$ and the full decomposition can be built up in a similar way but with the added complication of charge fractionalisation. 

{\color{black} Of course} the magnetic moments in spin ice are not needles. In real materials the magnetic fields spread out over all space \cite{Sala12a} giving corrections to this model that include ordered phases \cite{denHertog00,McClarty15a,Henelius16a}, corrections to diffuse neutron scattering predictions \cite{Sen13a,twengstrom19a} and to measurements from local probes \cite{Bramwell09,Dunsiger11a,chang13a}. However, as we show below it has proved extremely successful in predicting spin correlations in the partially ordered monopole and magnetic charge crystal phases discussed below.



\subsection{Coexistence of order and disorder}

The decomposition can be dealt with easily in the lower symmetry situation of a system driven into a monopole crystal phase by varying the chemical potential \cite{Brooks14,Borzi13}. Since magnetic monopoles interact via a Coulomb potential (see Eq.~(\ref{eq2})), they can form a stable structure of alternating positive and negative charges on the diamond lattice (see Fig.~\ref{fig:frag}); this is the well known zinc blende structure of ZnS. This structure appears naturally by tuning the chemical potential and forbidding double charges \cite{Brooks14,Borzi13} or in a staggered potential \cite{Lefrancois17,Raban19}. In other words the moments break up alternatively into the distribution exposed in Eq.~(\ref{eq:frag}) {\color{black} and its opposite}. Hence while the divergence-full field $\vec M_{\mathrm{m}}$ is long-range ordered, the divergence-free gauge field persists and forms a fluctuating magnetic background with characteristics of the Coulomb phase. The local degree of freedom of this fluctuating phase is the position of the minority spin in each tetrahedron, corresponding to the maximum flux $\pm 3/2$ exemplified in the right term of Eq.~(\ref{eq:frag}). As such, this particular Coulomb phase is exactly mapped onto a hard-core dimer model on the diamond lattice (see Fig.~\ref{fig:frag}) whose number of configurations is $\sim 1.3^{N/2}$ \cite{nagle66c} where $N$ is the number of pyrochlore sites.

As mentioned previously, there are also several mechanisms to stabilise such a phase made of alternating positive and negative charges ($\pm Q/2$) in two dimensions. On one hand, this can be done by dimensional reduction using a [111] magnetic field in three-dimensional spin-ice systems (see section \ref{sec:dimred}). It is the so-called kagome-ice plateau \cite{Matsuhira02b,Moessner03,Lhotel18} where the $\mathbb{Z}_{2}$ time-reversal symmetry of the long-range order component is intrinsically broken by the field. On the other hand, this $\mathbb{Z}_{2}$ symmetry can also be spontaneously broken in either dipolar kagome ice, or the kagome dumbbell model in which the long range part of the dipolar interaction appears as Coulomb interactions between emergent monopoles, forming the kagome-ice phase  (see section \ref{sec:ASI}). In both cases, using Eq.~(\ref{eq:fragk}), any spin configuration can be fragmented into a monopole crystal with charges sitting on the honeycomb sites (formed by the centres of the triangles on kagome) and a hard-core dimer model on the honeycomb bonds with residual entropy $S= 0.108\,k_{B}$ \cite{Moller09} (see Fig.~\ref{fig:fragk}). Explicitly, this means that the KI 2 phase has unsaturated antiferromagnetic order (the charge crystal) coexisting with the fluctuating Coulomb phase. Note that in both 2D and 3D, corrections beyond the dumbbell model, i.e. including the full dipolar interactions, ultimately order the system at very low temperatures  \cite{Moller09,Chern11,Borzi13,Jaubert15b}.

\begin{figure} [ht]
\centering\includegraphics[width=12cm]{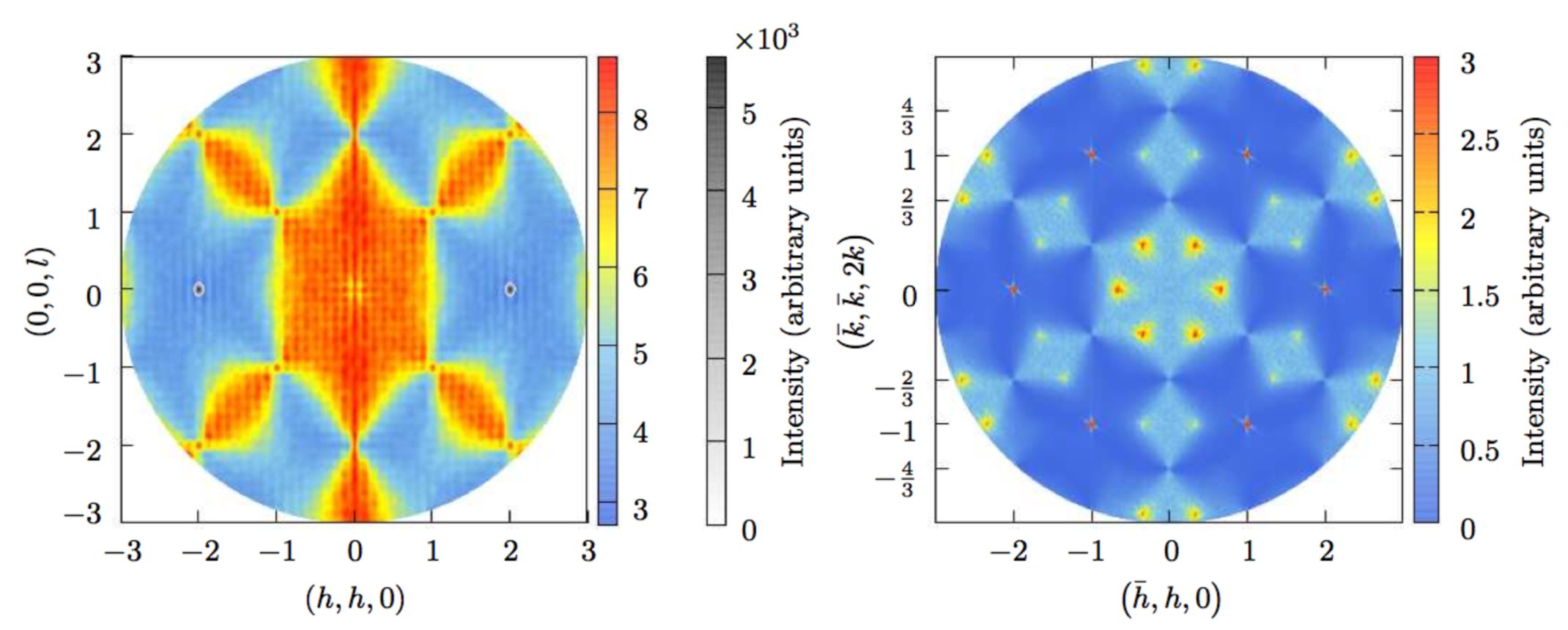}
\caption[Structure factor of a fragmented spin liquid]{\textbf{Structure factor of a fragmented spin liquid:} Simulated magnetic scattering function, $S(Q)$ for the pyrochlore monopole crystal (left) and for in-plane scattering from kagome ice (right). To reveal the diffuse scattering the Bragg peaks in the  pyrochlore data are plotted as contours in grayscale superimposed on the contribution to $S(Q)$. Adapted from Ref.~\cite{Brooks14}.
\label{fig:fragSQ}}
\end{figure}

This is a spectacular situation if one is looking for novel magnetic phases beyond the traditional picture of spin liquids, as the fragmentation results in the coexistence of long range {\color{black} order with a fluctuating gauge field [Figs.~\ref{fig:frag} and \ref{fig:fragk}]. In the canonical examples presented so far, the magnetic order is a crystal of monopoles while  the gauge field corresponds to a Coulomb spin liquid with pinch point scattering pattern. However, fragmentation could also apply to a broader diversity of ordered phases and emergent gauge fields, such as tensor gauge fields for example  \cite{benton16a,prem18a,yan20a}}. Simulated neutron scattering patterns are shown in Fig.~\ref{fig:fragSQ} for the above-mentioned monopole crystal in 3D and for the so called KI 2 phase of the analogous two-dimensional kagome ice (see section \ref{sec:ASI}). The figures show clear evidence of both Bragg peaks and diffuse scattering in a pinch point pattern. For convenience, we shall refer to this phase as a fragmented spin liquid (FSL).

Fragmentation should be distinguished from other concepts where order and disorder co-exist. \textcolor{black}{In particular, the long-range order and fluctuating spin liquid do not phase separate in different domains of the system, but co-exist everywhere via the fragmentation of the microscopic degrees of freedom (see Eq.~(\ref{eq:frag})).} {\color{black}Coexistence of short-range and long-range orders, often associated to different components of the magnetic moments have been reported in numerous low dimensional and/or frustrated systems (see e.g. \cite{Petrenko14} or \cite{Stewart04, Lefrancois19} in the Gd pyrochlore systems). Another scenario is the reduction of the ordered magnetic moment, together with the persistence of quantum spin fluctuations down to the lowest temperatures. This is often related to the proximity to a competing phase, or a quantum critical point. In pyrochlores, it is for example the case of Yb$_2$Ti$_2$O$_7$ \cite{Jaubert15, Robert15, Scheie19} or Er$_2$Sn$_2$O$_7$ \cite{Petit17,Yan17}.} \textcolor{black}{However, fragmentation differs from traditional quantum melting since its key point is that the disordered contribution supports an emergent gauge field.}

{\color{black} While the concepts} presented so far are classical, the notion of fragmentation can be extended to quantum systems \cite{savary12a,savary13a,Petit16b,Benton16b}\textcolor{black}{, for example in relation with quantum kagome ice where a field-induced finite magnetisation may co-exist with a $\mathbb{Z}_{2}$ spin liquid \cite{Carrasquilla15,Huang17,Wu19,Wang20a}. Fragmentation can also occur via the decoupling of the equations of spin motions. This has been proposed as an elegant mechanism for Nd$_2$Zr$_2$O$_7$ (see section \ref{sec:Nd}) whose inelastic magnetic structure factor fragments into flat bands with divergenceless fluctuations and divergence-full fluctuations forming Bragg peaks and dispersive bands \cite{Benton16b}. For a variety of frustrated systems \cite{Yan18,Mizoguchi18}, the divergence-full fluctuations also form pinch-point patterns but on the dispersive band of the energy spectrum. Measurements at iso-energy thus provide a cut of these dispersive pinch points in the characteristic form of half-moons \cite{Yan18,Mizoguchi18}.}

The fragmented fluid presented \textcolor{black}{in this section} is actually the classical version of the quantum dimer model on the diamond lattice \cite{Bergman06a,Sikora09a,Sikora11a} (modulo the time-reversal broken symmetry), which has been discussed in the context of the magnetisation plateau observed in HgCr$_{2}$O$_{4}$ and CdCr$_{2}$O$_{4}$ \cite{Penc04a,Ueda05a,Bergman06b}. It is also related to the quantum Coulomb ferromagnet of the extended quantum spin ice Hamiltonian \cite{ross11a,savary12a,savary13a}. In the broad picture, fragmentation is kindred to a family of exotic spin liquids with broken symmetry, such as chiral spin liquids \textit{\`a la} Kalmeyer-Laughlin \cite{Kalmeyer87a,Kalmeyer89a}, nematic spin liquids \cite{Grover10a} or even spin-lattice liquid with loop-length symmetry breaking \cite{Smerald19a}...

\subsection{Phase diagram of the fragmented spin liquid}
\label{sec:phasediag}
The Coulomb energy of the monopole crystal is
\begin{eqnarray}
U_C\;=\; -\; \frac{N_0}{2}\;\alpha\;\frac{\mu_0 Q^2}{4\pi a}
\label{eq:madelung}
\end{eqnarray}
where $N_0$ is the number of diamond sites and $\alpha$ is the Madelung constant. In three dimensions for the diamond lattice, it is $\alpha=1.638$. In the absence of double monopoles the monopole crystal would  become the ground state for \cite{Brooks14}
%
\begin{equation}
\mu > \mu^{\ast} = -\; \frac{\alpha}{2}\;\frac{\mu_0 Q^2}{4\pi a}.
\label{eq:muast}
\end{equation}
However, in the derivation from spin ice, $\mu_2=4\mu$, while the Coulomb energy also scales by a factor of four, 
so that a double monopole crystal, the ``all-in-all-out'' (AIAO) antiferromagnetic phase occurs at the same threshold. This masks the FSL phase unless we impose the limit ($\mu_{2}\rightarrow -\infty$), excluding double monopoles from the system. 
Similarly, the monopole crystal can also be stabilised over a finite fraction of the system when working at fixed concentration of single charges \cite{Borzi13,guruciaga14a}. Introduction of a four-body interaction \cite{Jaubert15b} lifts the constraint between $\mu$ and $\mu_2$, allowing the realisation of the FSL over a finite range of parameter space. Although this term is difficult to realise in experiments it could mimic the effects of quantum fluctuations as discussed further below.

\begin{figure}[t]
\centering\includegraphics[width=8cm]{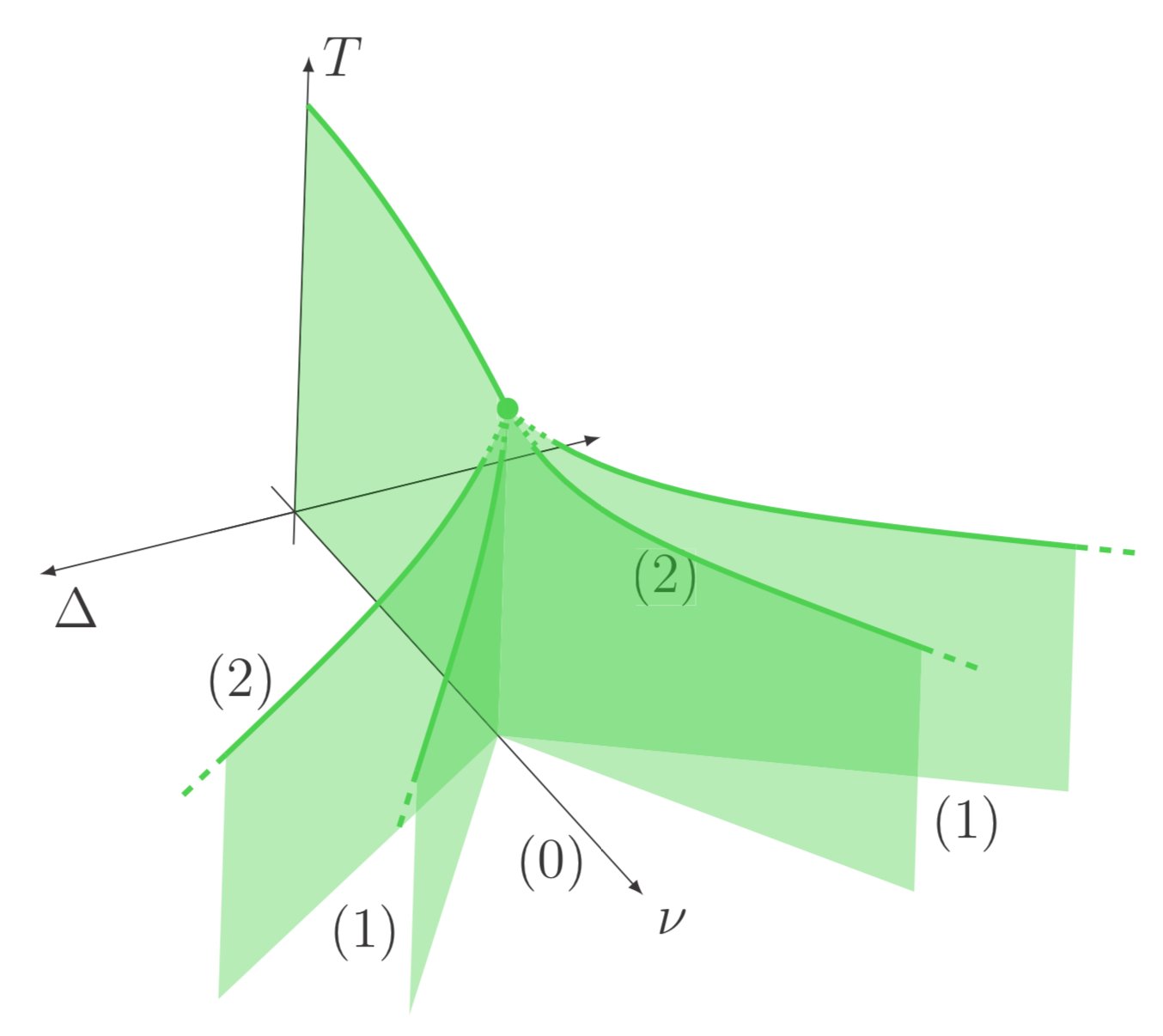}
\caption{\textbf{Phase diagram of the fragmented spin liquid:}
(0) monopole fluid, (1) monopole crystal, (2) double monopole crystal. Surfaces show $1^{st}$ and lines $2^{nd}$ order transitions. The central dotted lines show a multicritical region. The phase diagram corresponds to the dumbbell model with $\nu=-\mu=-\mu_2/4$ and a staggered potential $\Delta$. Reproduced from Ref.~\cite{Raban19}.}
\label{figPhD}
\end{figure} 



The AIAO and FSL phases can be separated in the dumbbell model by the application of a 
staggered potential $\Delta$ \cite{Brooks14,Raban19}, a scenario motivated in part by the action of Ir$^{4+}$ ions on Ho$^{3+}$ ions in Ho$_2$Ir$_2$O$_7$ \cite{Lefrancois17} (see section \ref{sec:ir}). This staggered potential breaks the \textcolor{black}{space group symmetry} of the underlying diamond lattice by favouring alternatively positive and negative charges on each of the bipartite sublattices, thus enforcing the symmetry of the zinc blend structure [Fig.~\ref{fig:frag}]. The resulting rich double winged phase diagram is shown in Fig. \ref{figPhD}.\\

The plane with $\Delta=0$ corresponds closely to the phase diagram of the DSI  above the small energy scale of the band of Pauling states \cite{Melko01,Melko04}, with the crystal of double monopoles (AIAO) separated from the spin ice monopole vacuum by a phase transition that passes from first  order at low tempareature to second order via a tri-critical point.  At the transition point, at zero temperature the double and single monopole crystal phases are degenerate but away from it the fragmented single monopole crystal phase is suppressed \cite{Brooks14}. Adding $\Delta$ separates the two phases giving the double wings; planes of first order and lines of second order transitions, separating five different phases \cite{Raban19}: the monopole fluid (spin ice physics with a non saturated density of monopoles), the double monopole crystal and the FSL phase. The ground states of the first two phases are extensively degenerate, while the last two phases have long-range order related by time-reversal symmetry when reversing the sign of $\Delta$ \cite{Cathelin20}. Indeed, as $\Delta$ breaks the \textcolor{black}{space group symmetry}, there is no further microscopic symmetry breaking and all transitions are symmetry sustaining. In this sense, the transition from monopole fluid to single monopole crystal is thermodynamically equivalent to the liquid-gas transition and that from single to double monopole crystal provides a skeleton for liquid-liquid phase transitions observed experimentally in supercooled liquids \cite{Poole92a,Katayama03a,Sastry03a,Brovchenko05a}. 
The transition from monopole fluid to FSL is a close cousin to the phase transition observed in spin ice in the presence of a $[111]$ field in which the system leaves the kagome ice plateau via a monopole crystallisation transition \cite{Matsuhira02b,Castelnovo08}. In that case, the external field couples to both the $\vec M_{\mathrm{d}}$ and $\vec M_{\mathrm{m}}$, whereas the staggered potential $\Delta$ couples to $\vec M_{\mathrm{m}}$  only \cite{Raban19}. 

This is the generic phase diagram of the $S=2$ Blume-Capel model \cite{Lara98} and analogous sets of phase transitions can be observed elsewhere, for example in itinerant magnetic compounds under pressure and in an external field \cite{Kaluarachchi17a,Kotegawa11a}. The form of the multi-criticality as the five different phases merge towards each other is non-universal, depending on microscopic parameters, but a single penta-critical point is not expected \cite{Raban19}.  

\subsection{Excitations of a fragmented spin liquid and monopole holes}
\label{sec:excitations}

In the traditional spin ice model, the gauge charges are the 3 in - 1 out and 3 out - 1 in tetrahedron states, corresponding to single-charged magnetic monopoles in the dumbbell model. 
In the ground state of the fragmented spin liquid, tetrahedra are alternatively 3 in - 1 out and 3 out - 1 in. Excitations out of the FSL are thus 4 in / 4 out (also called all in / all out) and 2 in - 2 out states. These excitations have a dual representation, either in terms of the original emergent fields, or as gauge charges out of a new emergent field, which is the transverse fragment of the spin ice field $\vec M_{\rm d}$. The excitations will create a new longitudinal field $\vec {\tilde{M}}_{\rm m}$ which adds vectorially to the total transverse component, $\vec M_{\rm m}$.

This duality gives a crucial difference between spin-ice and FSL charges. In the former case, there is a perfect symmetry between positive and negative single charges -tetrahedra with 3 in - 1 out and 3 out - 1 in. In the latter case however, single gauge charges that give 2 in - 2 out or 4 in / 4 out tetradedra have a priori different chemical potentials. And the picture gets more complicated for higher charges \cite{Jaubert15b}. 
Hence, while in spin ice creation of gauge charges can only increase the contribution of the divergence-full field $\vec M_{\rm m}$, in the FSL the total longitudinal contribution can either increase, as in the case of a 4 in or 4 out defect, or decrease if the move is to a 2 in - 2 out tetrahedron. For this second class of excitation, the constraint (\ref{constraint-2}) gives a corresponding increase in the divergence-free field $\vec M_{\rm d}$. In either case, the total contribution from $\vec M_{\rm m}$ or of $\vec {\tilde{M}}_{\rm m}$ is minimised if like charges are separated by a maximum distance and unlike charges are drawn together, generating effective Coulomb interactions between the gauge charges.

These ideas can be tested both for the NNSI and for the DSI.
In the nearest-neighbour spin-ice model, i.e. in absence of dipolar interactions, excitations out of the Coulomb phase lose their magnetic charges as we have discussed above, but remain gauge charges of the emergent field with an entropic Coulomb potential \cite{Henley10}. The entropic Coulomb potential comes from the ensuing configurational entropy difference between states with gauge particles separated by different distances.
\begin{eqnarray}
E^{\rm ent}=\rho\frac{k_{B} T}{R},
\label{eq:entropicV}
\end{eqnarray}
where $T$ is the temperature and $R=r/a$ is the dimensionless distance between gauge charges. $\rho$ is a characteristic parameter of the Coulomb phase, proportional to its stiffness~\cite{Henley10,Castelnovo11a}. Eq.~(\ref{eq:entropicV}) results in a temperature-independent probability
\begin{eqnarray}
P(\vec R)\propto \exp\left(E^{\rm ent}/k_{B} T\right) = \exp\left(\rho/R\right).
\label{eq:entropicP}
\end{eqnarray}
for two gauge charges to be at distance $R$.

\begin{figure}[t]
\centering\includegraphics[width=8cm]{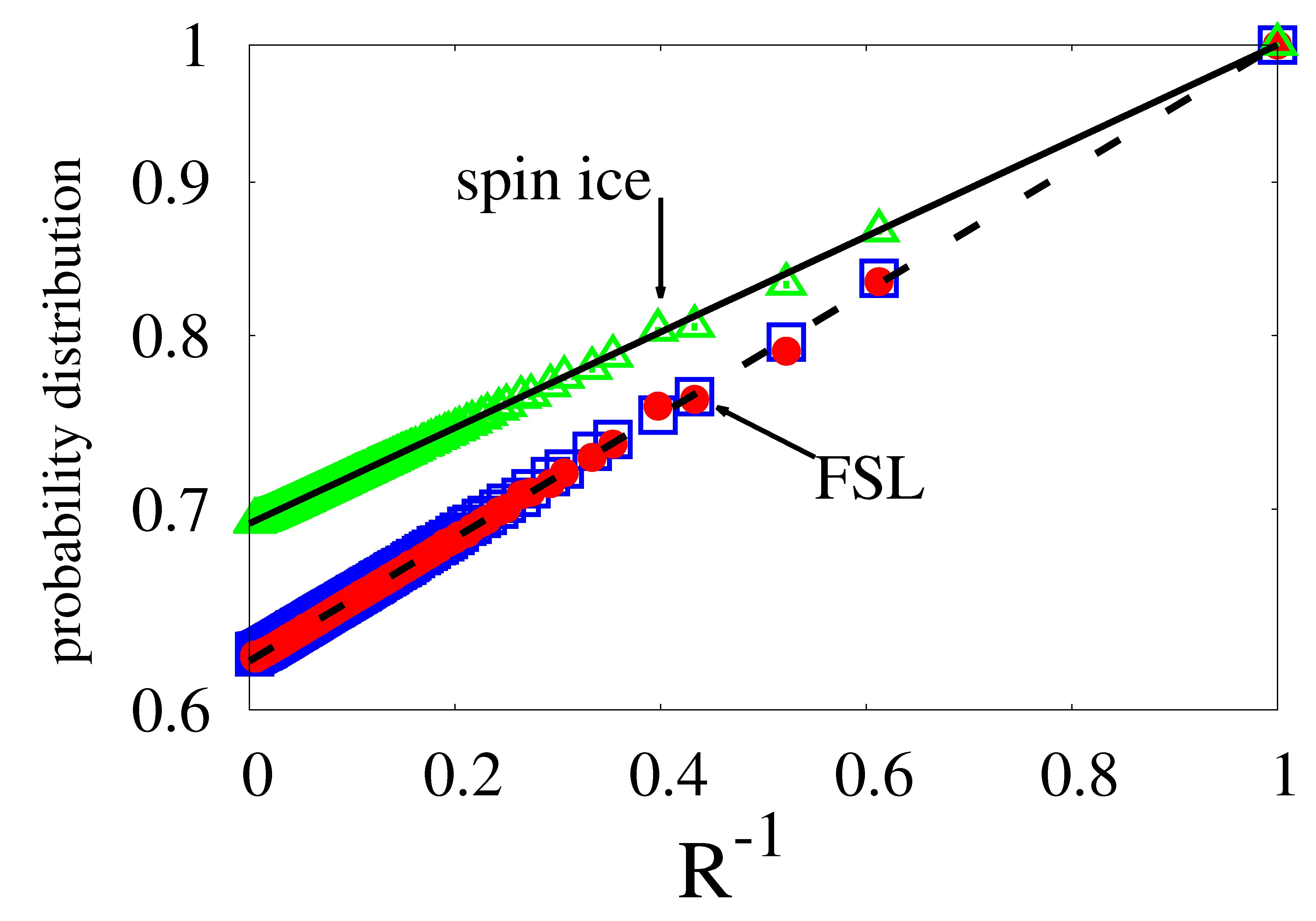}
\caption{
Monte Carlo simulations confirm the entropic probability distribution (\ref{eq:entropicP}) for a unique pair of gauge charges to be separated by a distance $R$, out of the 2 in - 2 out spin-ice ground state (\textcolor{green}{$\triangle$}) and out of the FSL (\textcolor{blue}{$\square$} and \textcolor{red}{$\bullet$}). For the sake of completeness, the two cases where the initial pair of defects out of the FCSL were 2 in - 2 out (\textcolor{red}{$\bullet$}) and 4in-4out (\textcolor{blue}{$\square$}) are shown to give the same outcome with double-spin dynamics. The $y-$axis is on a logarithmic scale, while the $x-$axis is on a linear scale. The error bars are smaller than the data symbols. Adapted from Ref.~\cite{Jaubert15b}.
}
\label{fig:entropicP}
\end{figure}

The probability of Eq.~(\ref{eq:entropicP}) has been measured in Monte Carlo simulations of the spin-ice ground-state \cite{Castelnovo11a,Jaubert15b} and of the FSL \cite{Jaubert15b}, confirming the presence of the entropic potential (\ref{eq:entropicV}) with $\rho_{\rm spin-ice}=0.36755$ and $\rho_{\rm FSL}=0.473$ (see Fig.~\ref{fig:entropicP}). Moving to the DSI, it was shown that with the inclusion of the magnetic flux, the gauge charges out of the FSL transform into magnetic charges with an effective Coulomb force between them \cite{Jaubert15b}.

Spin ice has often been described as a ``magnetolyte'' \cite{Bramwell09,Giblin11,zhou11a,Castelnovo11a} with essentially similar positive and negative charges emerging from a neutral solution. For the FSL, the analogy is closer to a semiconductor \cite{Jaubert15b}, where we have a valence band filled with electrons and an empty conduction band. Gapped excitations take the form of a pair of electric charges of opposite sign: an electron occupying the conduction band; and an electron hole occupying the valence band. Similarly in the FSL, we have co-existence of long-range order filled with magnetic monopoles, and a fluctuating spin liquid empty of charges. Excitations are also gapped and take the form of a pair of magnetic charges of opposite sign: a monopole, locally destroying the divergence-free field $\vec M_{\rm d}$ (as in spin ice); and a monopole hole, locally destroying the long-range order of the divergence-full field $\vec M_{\rm m}$. The effective Coulomb potential also ensures deconfinement of the charges and thus fractionalization of the excitations, as in semiconductors.

To conclude this discussion, the nature of these excitations -- monopoles and monopole holes -- is dynamically conserved via a double-spin motion, i.e.~ by flipping two neighbouring spins at the same time \cite{Jaubert15b,Lefrancois17}. This is due to the $\mathbb{Z}_{2}$ broken symmetry of the ordered phase making the divergence-full field alternatively a source and a sink of fluxes on the bipartite diamond lattice. This double-spin motion corresponds to a dimer move in the equivalent dimer model of the divergence-free fluid, and is reminiscent of the mobility of holons and spinons in frustrated Mott insulators on bipartite lattices~\cite{Poilblanc11a}. The double-spin motion is necessary to recover the effective Coulomb potential between magnetic charges \cite{Jaubert15b}. On the other hand, when a single-spin flip dynamics is considered, the $\mathbb{Z}_{2}$ broken symmetry then appears as a staggered potential \cite{Lefrancois17}, as explained in section \ref{sec:ir}.

\section{Experimental realisations}
As depicted in the previous section, an important feature of the fragmented phases is the coexistence of a Coulomb phase and an ordered phase. This manifests in the magnetic scattering function, which can be probed by neutron diffraction measurements, as the superposition of Bragg peaks with anisotropic diffuse scattering which exhibits pinch points, in the proportions predicted by the Helmholtz decomposition. Another important signature is the existence of a residual entropy despite the presence of a phase transition.

In the following we address the two lattices in which experimental fragmented phases have been reported, the two dimensional kagome lattice and the three dimensional pyrochlore lattice. 

\subsection{Magnetic fragmentation in kagome systems}
Systems which are good candidates to stabilize fragmented phases are kagome ice systems. 
Indeed, as detailed in Section \ref{sec:Helmholtz}, their ground state (2 in - 1 out / 1 in - 2 out configurations) is intrinsically charged. In the presence of nearest neighbor interactions only, these charges remain disordered down to zero temperature. But in the presence of dipolar interactions, charge ordering is predicted when the temperature is decreased, before entering a fully ordered phase, where both spins and charges are ordered \cite{Chern11}.  As mentioned previously, the intermediate phase, called KI 2, is nothing but a fragmented phase. 

\subsubsection{Artificial spin ice}
\label{sec:ASI}

Artificial spin ice systems, made of lithographically built arrays of nanomagnets interacting through dipolar interactions, are ideal to visualize in real space the magnetic arrangement and thus the presence of charge ordering. The main difficulty that arises experimentally is to achieve the thermodynamic ground state of the built lattice, because the system is essentially static at room temperature. Intense research has been dedicated to the development of experimental protocols or novel materials to be able to manipulate the magnetic configurations at room temperature and obtain magnetic states as close as possible to the thermodynamic ground state \cite{skjaervo19a,Rougemaille19}.   

\begin{figure}
\begin{center}
\includegraphics[height=5cm]{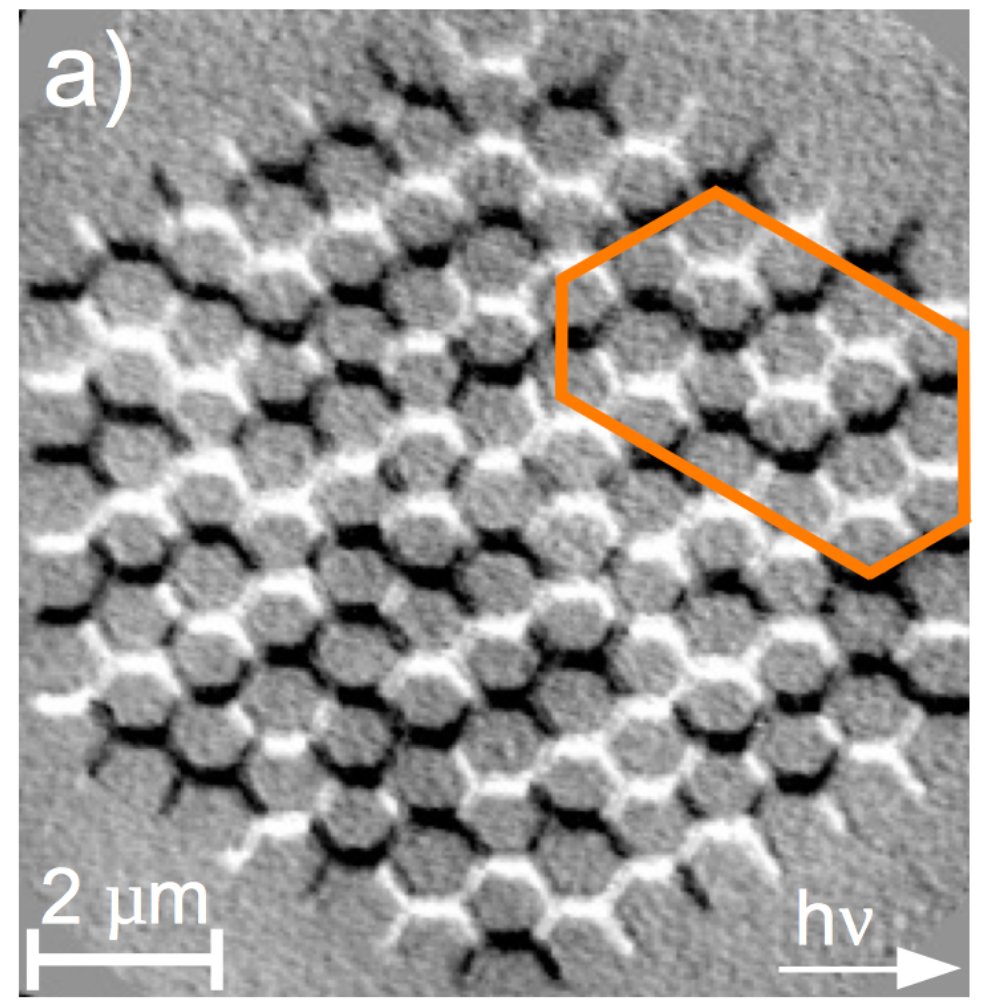}\includegraphics[height=5cm]{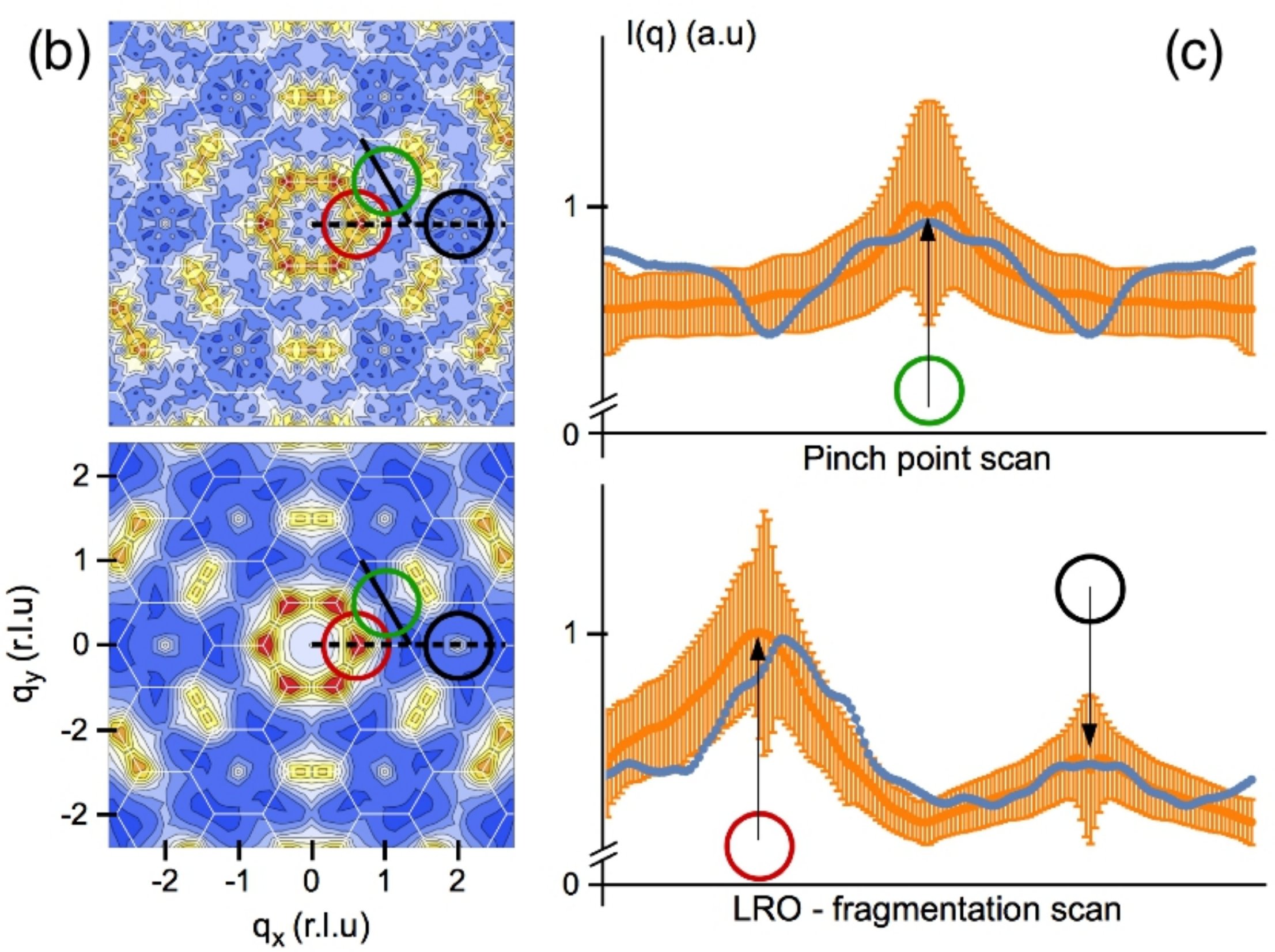}
\end{center}
\caption{(a) XMCD-PEEM image of an artificial  Gd$_{0.3}$Co$_{0.7}$ kagome array. The orange region corresponds to a charge ordered domain. (b) Experimental (top) and theroretical (bottom) maps of the magnetic correlation function in the reciprocal space corresponding the picture (a). (c) Corresponding $q$-cuts along the full (top) and dashed (bottom) lines, where the positions of the pinch point (top) and the Bragg peaks (bottom) are marked by the colored circles. Experimental signal (blue) is compared with theoretical calculations (orange). Adapted from \cite{Canals16, Rougemaille19}.  }
\label{ASI}    
\end{figure}

In this context, the charge ordering state could be achieved in kagome arrays of permalloy (Ni$_{80}$Fe$_{20}$) using an annealing protocol \cite{Zhang13}. Charge crystallites could be observed, while the magnetic configurations remain disordered. The residual entropy obtained from the analysis does not correspond to the fully ordered charge model, but is well below what is expected for the kagome ice phase only. The fragmented nature of this charge ordered state was evidenced by Canals et al. \cite{Canals16} in a Gd$_{0.3}$Co$_{0.7}$ array. The key is to show the coexistence of a Coulomb phase with the charge ordering. While artificial spin ices are often studied through the correlators in real space obtained from the magnetic pictures, the trick here was to Fourier transform these correlations into the magnetic scattering function in reciprocal space $S(Q)$, in order to visualize the pinch points, signature of the Coulomb phase and peaks, signature of the ordered phase (as previously illustrated in Figure \ref{fig:fragSQ}). The figure \ref{ASI} shows the real space picture (a), and the associated $S(Q)$ (b). A comparison with Monte-Carlo calculations could then show that the magnetic state of the GdCo array is very close to the fragmented phase, with the emergence of Bragg peaks above the kagome ice magnetic scattering function (See Figure \ref{ASI}(b) and (c)). 

\subsubsection{Tripod kagome}
\label{tripod}

\begin{figure}[b]
\begin{center}
\includegraphics[height=5cm]{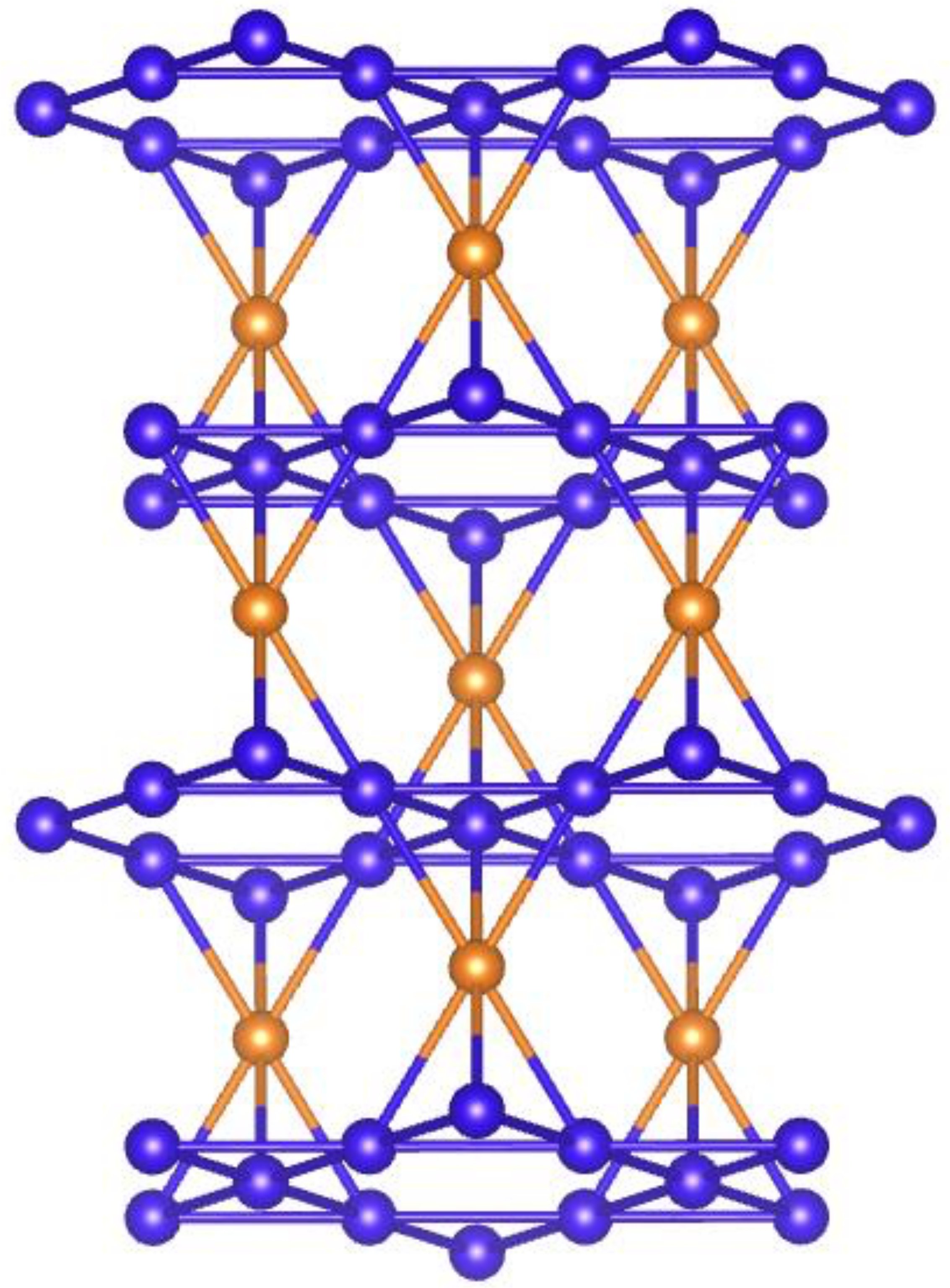} \qquad
\includegraphics[height=5cm]{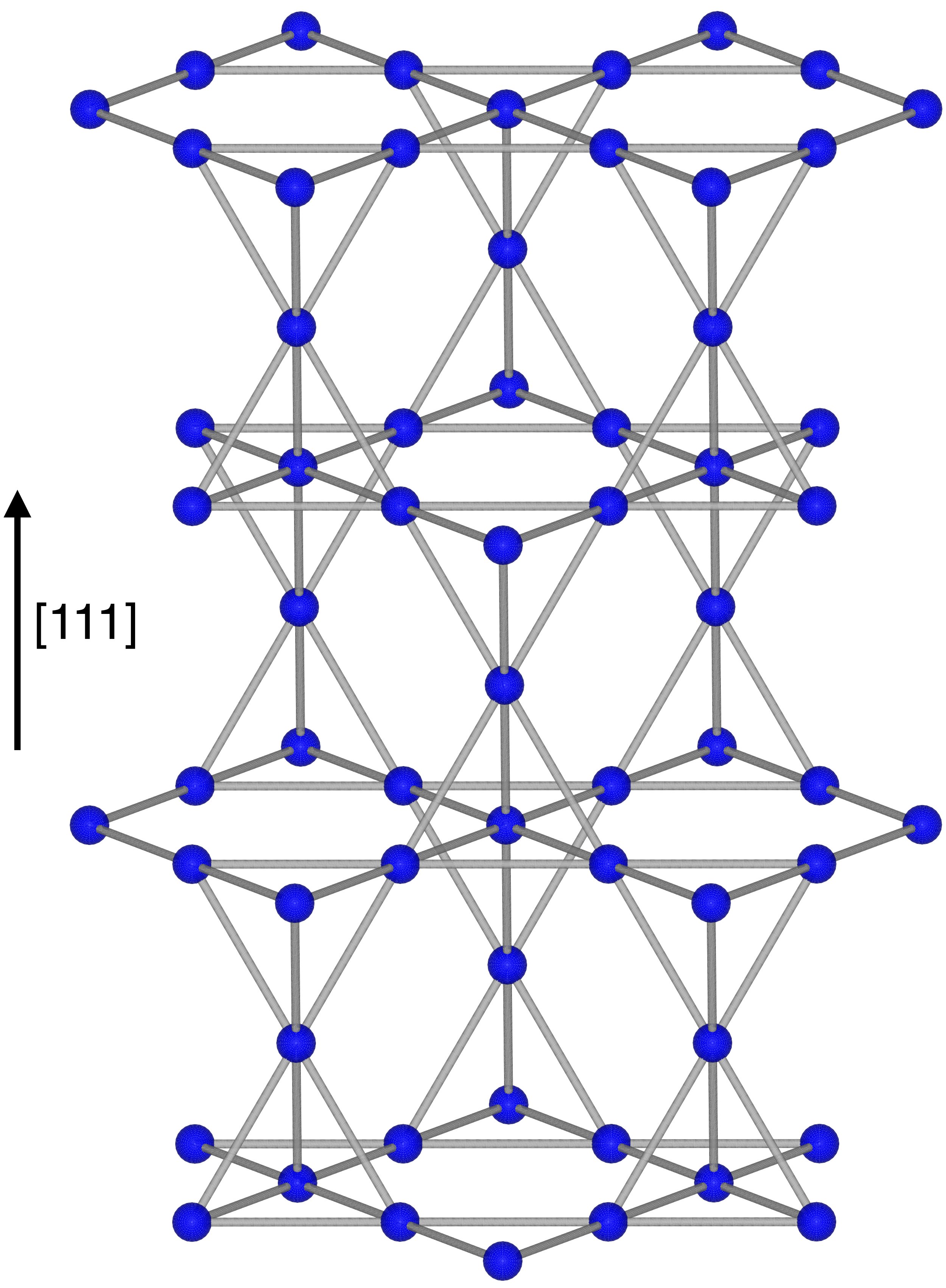}
\end{center}
\caption{View of the tripod kagome lattice (left)  and of the pyrochlore lattice (right) along the $[001]$ and $[111]$ directions respectively, showing the stacking of triangular and kagome planes. In the $R_3$Mg$_2$Sb$_3$O$_{14}$ compounds, the kagome planes are occupied by the rare-earth $R$ (blue) and the triangular planes are occupied by Mg (orange). }
\label{pyrochlore_111}    
\end{figure}

Concomitantly, emergent charge ordering was also reported in a bulk kagome system, Dy$_3$Mg$_2$Sb$_3$O$_{14}$ \cite{Paddison16}. This compound has actually a pyrochlore structure, but in which the triangular planes are occupied by non magnetic Mg, the magnetic Dy atoms thus occupying two dimensional kagome planes (See Figure \ref{pyrochlore_111}) \cite{Dun16, Dun17}. The environment of the Dy$^{3+}$ ions is pretty similar to pyrochlore systems. Their magnetic moments can thus be considered as Ising spins. They are not confined in the kagome planes but have a component out of the planes, which slightly affects the conventional kagome ice picture. The effective interaction, resulting from exchange and dipolar interactions, is ferromagnetic, making this compound an ideal candidate for magnetic fragmentation. Using neutron diffraction measurements, Paddison et al. \cite{Paddison16} could indeed show that, below 300 mK, diffuse scattering coexists with an all in / all out ordering, with a reduced moment of  2.8 $\mu_{\rm B}$ close to the expected value for the Dy$^{3+}$ ion in the fragmented kagome phase $\mu/3 \approx 3.3~\mu_{\rm B}$. Measurements were perfomed on a powder sample, which prevents the observation of the pinch points characteristic of the Coulomb phase. Reverse Monte-Carlo calculations could nevertheless show that the diffuse signal does correspond to Coulomb phase correlations. A small residual entropy was obtained from specific heat measurements ($\Delta=0.05 R$) and matches within error bars the expected $0.11 R$ value. The presence of Mg - Dy site disorder (estimated to about 6\% in the studied sample) is invoked to explain some discrepancies with the model. The authors also note that the model Hamiltonian they deduce from their experiments predicts a phase transition towards a full long-range order, that is not observed, maybe due to the presence of disorder or to frozen single spin-flip dynamics.   

The physics of Dy and Ho systems being very similar in pyrochlore oxides $R_2M_2$O$_7$ ($R=$ rare-earth, $M=$ metal), it is natural to think of the Ho member of this family as a possible candidate for the stabilization of a fragmented kagome ice state. 
In Ho$_3$Mg$_2$Sb$_3$O$_{14}$ powder samples, the coexistence of partial all in / all out ordering with diffuse scattering akin to that expected for a Coulomb phase has indeed been observed \cite{Dun18}. Nevertheless, the ordered magnetic moment is smaller than expected, (1.7  instead of 3.3 $\mu_{\rm B}$) and than in the Dy compound. In addition, low energy excitations are observed in inelastic neutron scattering measurements, contrary to the expectation for a conventional Ising spin, showing that the system remains dynamic.  
The main difference between both systems is that Dy$^{3+}$ is a Kramers ion, which guarantees a magnetic doublet ground state, while Ho$^{3+}$ is a non-Kramers ion. In pyrochlore oxides, the symmetry of the rare-earth site guarantees that the magnetic ground state of the rare-earth is a magnetic doublet, thus conferring almost the same properties to Dy and Ho compounds. In $R_3$Mg$_2$Sb$_3$O$_{14}$ systems, the symmetry of the rare-earth site is lower than in the pyrochlore lattice, thus lifting the degeneracy of the ground doublet for the non-Kramers ions such as Ho. Due to the small size of the crystal field splitting, Ho$_3$Mg$_2$Sb$_3$O$_{14}$ was proposed to be an example of a kagome ice in a transverse field Ising model \cite{Wang68} coupled to a nuclear spin bath, resulting in a quantum spin fragmented state modified by quantum dynamics \cite{Dun18,Dun19a}. 

The Nd compounds, which stabilize a partial all in / all out ordering \cite{Scheie16, Dun17}, may also be potential candidates for fragmentation.  

\subsection{Magnetic fragmentation in pyrochlore systems}
The stabilization of a fragmented state in pyrochlore systems is more tricky. In the case of effective ferromagnetic nearest neighbor interactions, the ground state is the spin ice state, which is not charged \cite{Melko04}. A pocket of FSL could be generated by zero-point quantum fluctuations which would liberate the constraint $\mu_2=4\mu$ [Eq.~(\ref{eq:mu})] \cite{Brooks14,Raban19}. No real material is known up to date which satisfies these conditions. \textcolor{black}{However, alternative mechanisms have appeared stabilising the fragmented phase in this review}.

\begin{figure}[b]
\begin{center}
\includegraphics[height=5cm]{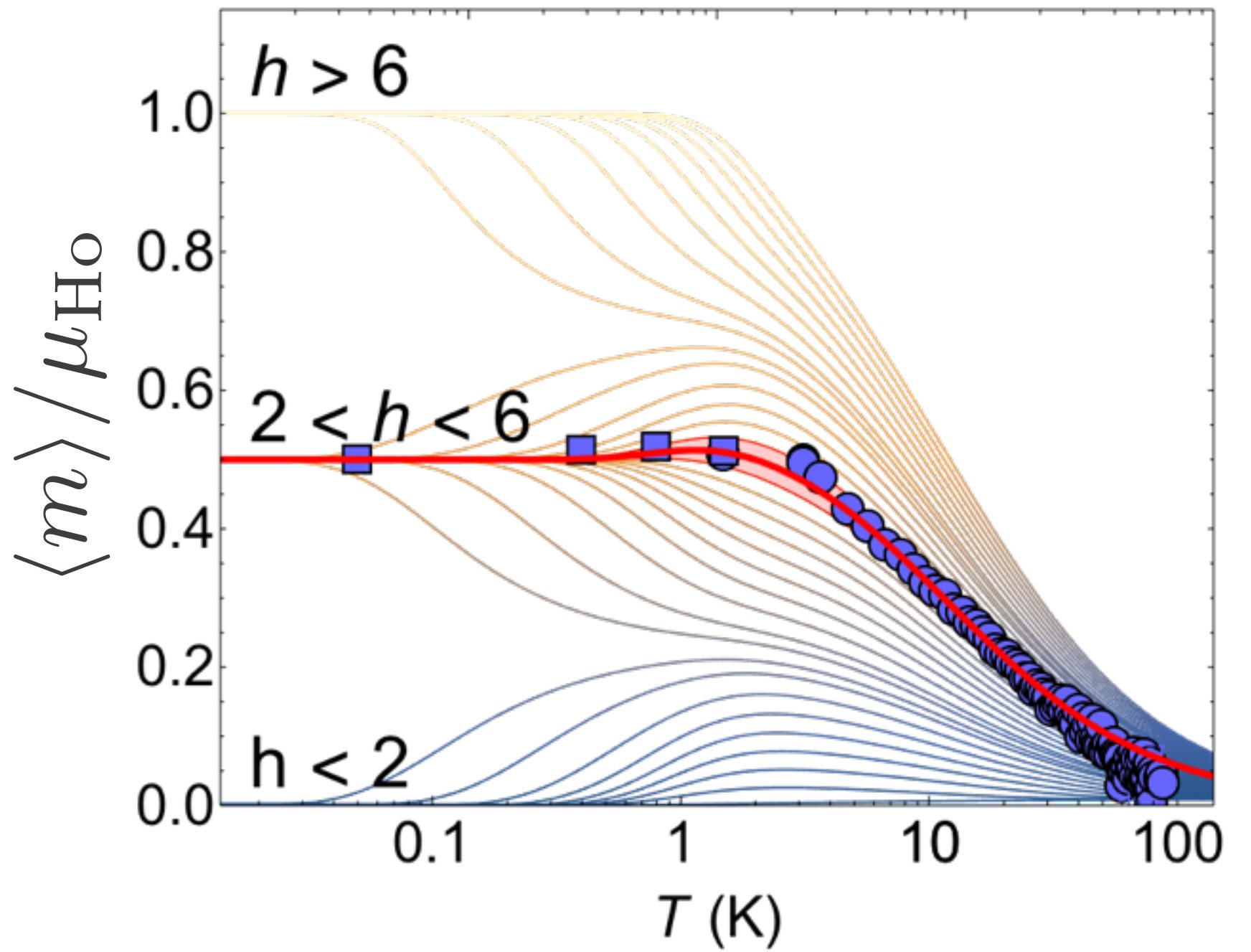} \quad
\includegraphics[height=5.5cm]{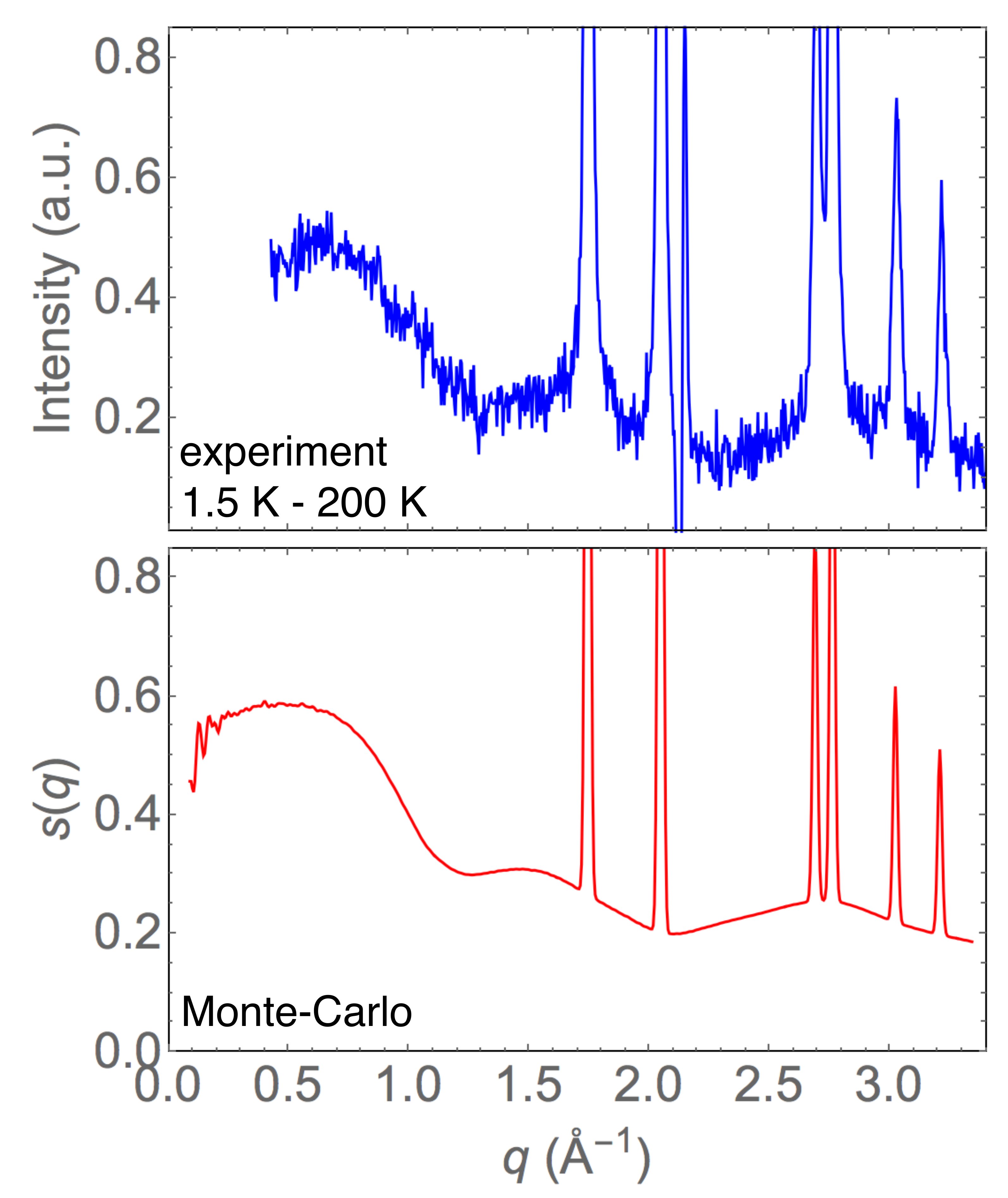}
\end{center}
\caption{left: Ordered all in / all out magnetic moment as a function of temperature obtained from neutron diffraction measurements (squares and dots), normalized to the ground doublet Ho magnetic moment. Lines are Monte-Carlo calculations in the NNSI model for different $h=h_{\rm loc}/J_{\rm eff}$ values. The red line shows the best agreement obtained for $h=4.5$. right: Diffuse scattering measured at 1.5 K (top) compared to the magnetic scattering function $S(q)$ calculated with Monte-Carlo calculations (bottom). Adapted from \cite{Lefrancois17}. }
\label{HoIr}    
\end{figure}

\subsubsection{Pyrochlore iridates}
\label{sec:ir}

Starting with a charge free spin ice ground state, another approach is to inject magnetic charges through an external parameter. This can be done through a staggered chemical potential as shown in Section \ref{sec:phasediag}. In materials, this staggered chemical potential is equivalent to a staggered magnetic field, aligned along the local $\langle 111 \rangle$ directions of the rare-earth sites. In such a scenario, the fragmented state can be stabilized providing the ratio between the effective interactions and the staggered field is large enough to favor magnetic charges (i.e. 3 in - 1 out / 3 out - 1 in configurations), but not too strong to prevent the stabilization of a double charge crystal (all in / all out) \cite{Lefrancois17}, as discussed above.  

This scenario can take place in pyrochlore iridates $R_2$Ir$_2$O$_7$. In these compounds (for all rare-earths except Pr), the magnetic Ir$^{4+}$ ions order antiferromagnetically at relatively high temperature (compared to the rare-earth interactions), between 30 and 150 K \cite{Matsuhira11}. The associated magnetic structure is all in / all out, which creates on the rare-earth ions a staggered molecular field $H^{\rm loc}$, precisely aligned along the local $\langle 111 \rangle$ directions \cite{Tomiyasu12, Lefrancois15}. The natural candidates that emerge are the Dy and Ho compounds, which are known to stabilize a spin ice ground state when the $M$ site is occupied by a non-magnetic ion \cite{Gardner10}. 

It was indeed shown that Ho$_2$Ir$_2$O$_7$ enters in a fragmented state below about 1 K \cite{Lefrancois17}. The key experimental features are obtained from neutron diffraction measurements: a partial ordering of the magnetic moment with a value equal to half of the Ho magnetic moment at low temperature, associated to diffuse scattering which persists down to the lowest temperature (See Figure \ref{HoIr}). No single crystals are available for this sample, but the powder averaging of the pinch point pattern expected for a single crystal matches the $q$ dependence of the measured diffuse scattering. The temperature dependence of the magnetic moment can be reproduced by a nearest neighbor spin ice (NNSI) model and gives a ratio $h_{\rm loc}/J_{\rm eff}=4.5$, where $J_{\rm eff}$ is the effective nearest neighbor interaction (resulting from the antiferromagnetic exchange and the dipolar interaction). This value places this compound deep in the fragmented phase, which, in this NNSI model, is stabilized for $2<h_{\rm loc}/J_{\rm eff}<6$. 

 It is worth noting that in this model, there is no need for long-range interactions to stabilize the charge crystal, since it is favored locally by the staggered molecular field (or chemical potential). Interestingly, these measurements could highlight the novel dynamics of the magnetic excitations that emerge from the fragmented phase compared to the canonical spin ice case, and discussed in Section \ref{sec:excitations} \cite{Jaubert15b}. Here, in the presence of the staggered chemical potential, the magnetic excitations can be viewed as charged excitations emerging from the divergence free phase, which propagate in a staggered potential created by the charge ordered part of the fragmented phase. This picture can be compared with the case of spin ice where the ground state is a vacuum of charges on which magnetic monopoles can propagate almost freely \cite{Castelnovo12}. Signatures of this dynamics are observed in the Ho$_2$Ir$_2$O$_7$ ac susceptibility, where the frequency dependence can be described by an Arrhenius law with a mean energy barrier resulting from the interaction between the charged excitations and the underlying staggered potential \cite{Lefrancois17}.    

More recently, it was shown that Dy$_2$Ir$_2$O$_7$ also stabilizes a fragmented phase, with almost the same characteristics as the Ho compound \cite{Cathelin20}. The specific heat measurements performed in this compound could confirm the existence of a residual entropy, in agreement with the theoretical predictions \cite{Raban19}. Further analysis could point out the importance of accounting for long-range interactions to quantitatively describe the fragmented phases of these iridates. In particular, calculations with the dumbbell model considerably improve the agreement with experiments, allowing with the same parameters for the reproduction of the temperature dependence of the magnetic moment and of the specific heat peak, as well as the excitation energy scale.

\subsubsection{Nd-based pyrochlores}
\label{sec:Nd}    

\begin{figure}[b]
\begin{center}
\includegraphics[height=4.3cm]{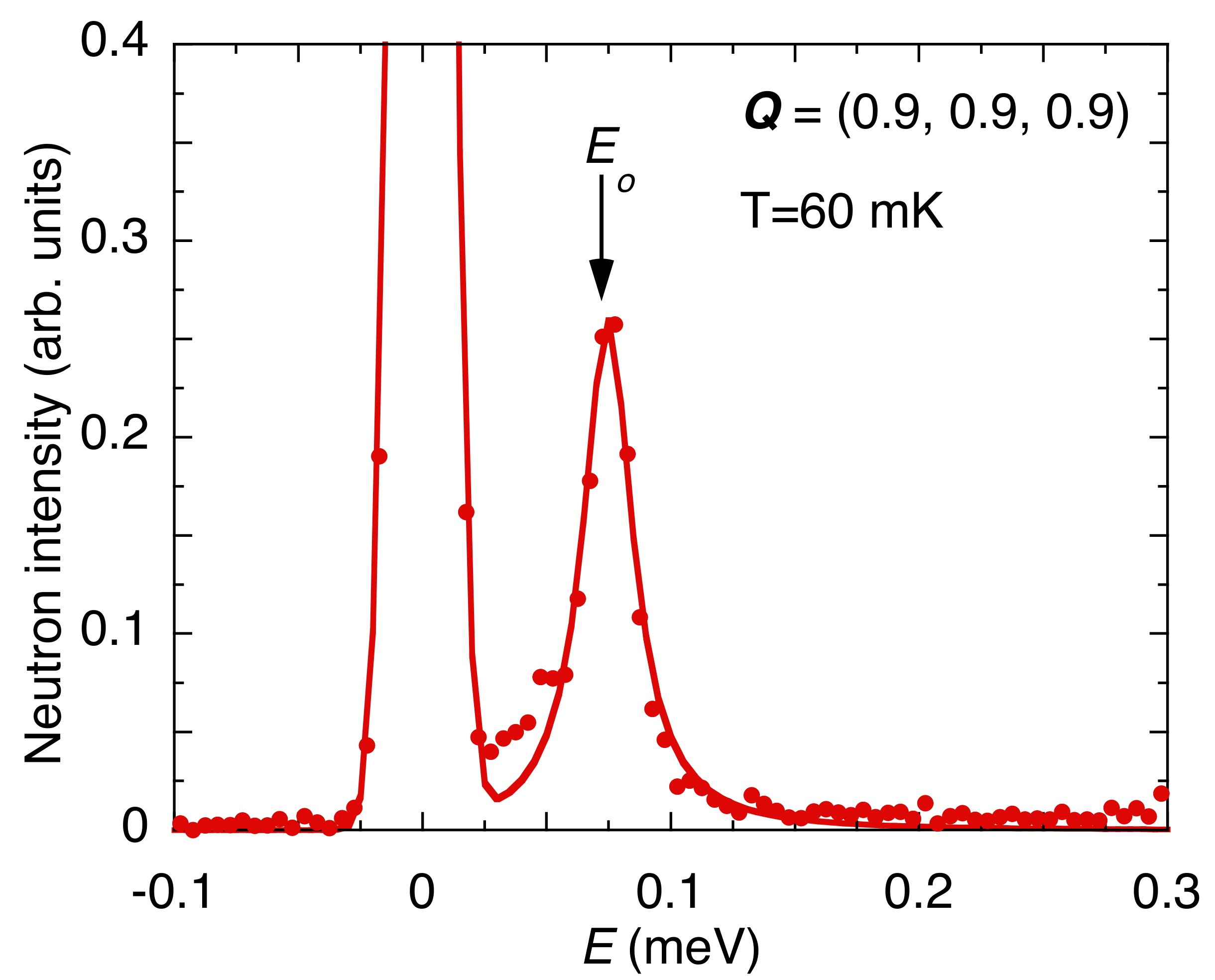}
\includegraphics[height=4.3cm]{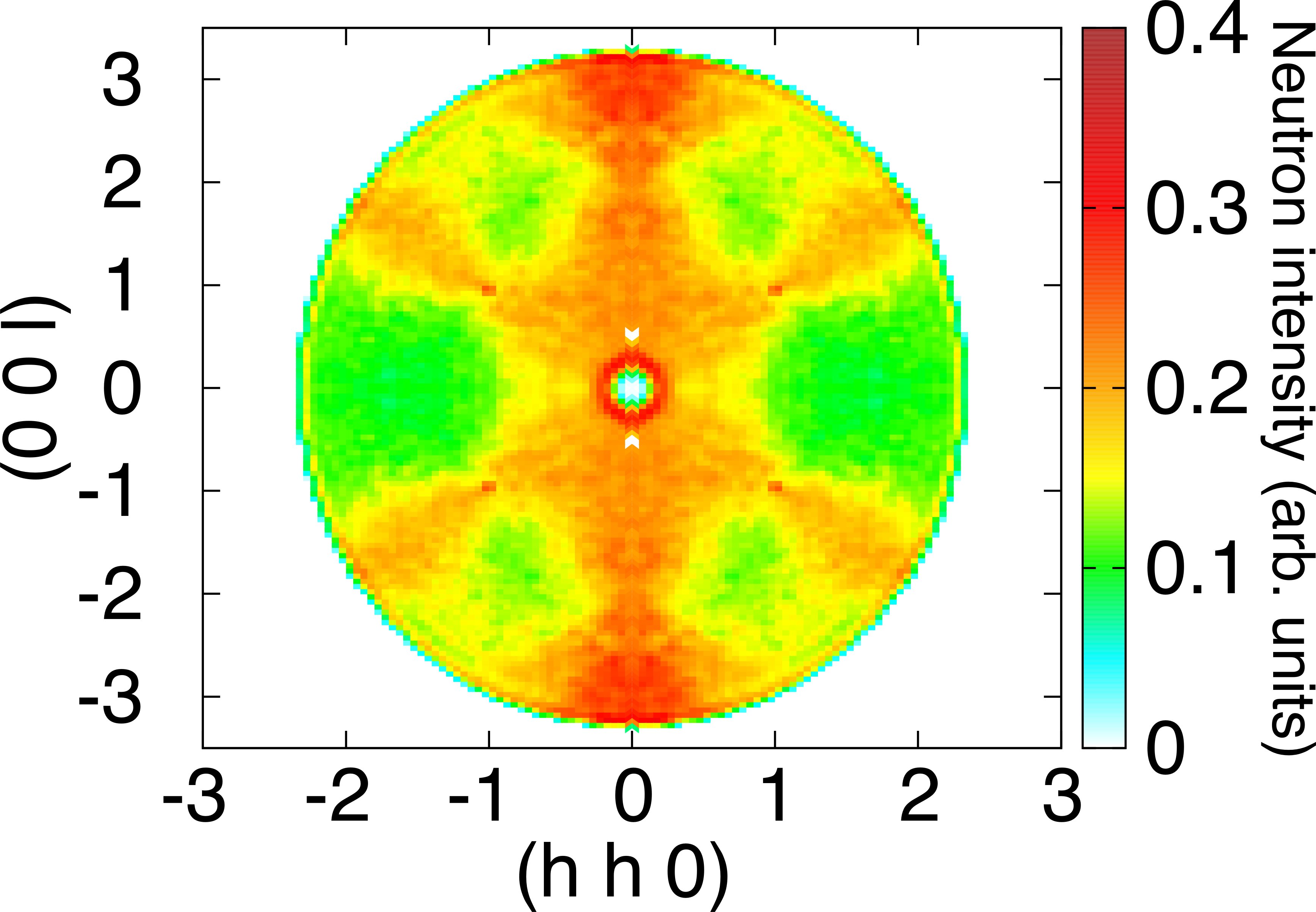} 
\end{center}
\caption{Inelastic neutron scattering in Nd$_2$Zr$_2$O$_7$ measured at 60 mK. left: Excitation spectrum at ${\bf q}=(0.9,0.9,0.9)$. right: Map averaged in the intensity range $45<E<55~\mu$eV, corresponding to the peak intensity in the left figure. Adapted from \cite{Petit16b}}
\label{NdZr_INS}    
\end{figure}

The first pyrochlore oxide study which refers to magnetic fragmentation concerns the Nd$_2$Zr$_2$O$_7$ compound. This system apparently presents all the ingredients for magnetic fragmentation: Nd$^{3+}$ magnetic moments are Ising-like, pointing along the local $\langle 111 \rangle$ directions, magnetic susceptibility exhibits a small but positive Curie-Weiss temperature, characteristic of effective interactions that are weakly ferromagnetic and a partial all in / all out ordering is observed \cite{Ciomaga15, Lhotel15, Xu15}. Nevertheless, the ordered magnetic moment is about one third of the total magnetic moment and is slightly sample dependent, which takes us away from the ``conventional" fragmentation scenario. A pinch point pattern, characteristic of the existence of a divergence free component, was observed in neutron scattering experiments. However, this signal is at finite energy \cite{Petit16b} (See Figure \ref{NdZr_INS}), which means that the divergence free contribution is associated with magnetic excitations and not to the ground state of the system. This is in contrast with the fragmentation scenario described in Section \ref{sec:theory}, where the Coulomb phase associated with the divergence free component is part of the ground state of the system, and thus is expected to give a signal at zero energy.  In addition, in Nd$_2$Zr$_2$O$_7$, above this spin ice pattern, dispersive branches emerge which indicates some propagation of the excitations. The related compound Nd$_2$Hf$_2$O$_7$ shows similar features \cite{Anand15}, while the stannate compound fully orders in an all in / all out state \cite{Bertin15}.   

\begin{figure}[b]
\begin{center}
\includegraphics[height=3cm]{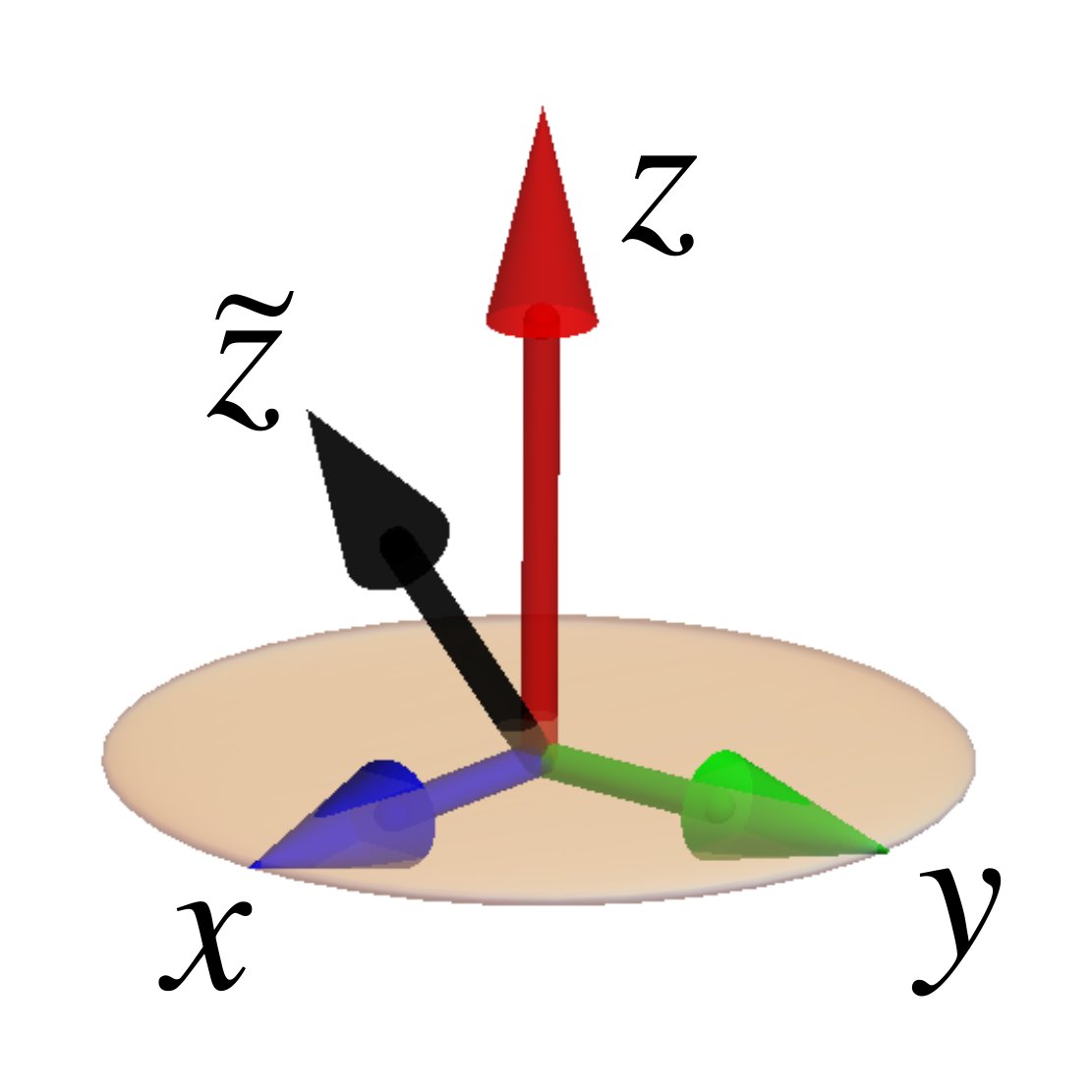} \qquad
\includegraphics[height=4cm]{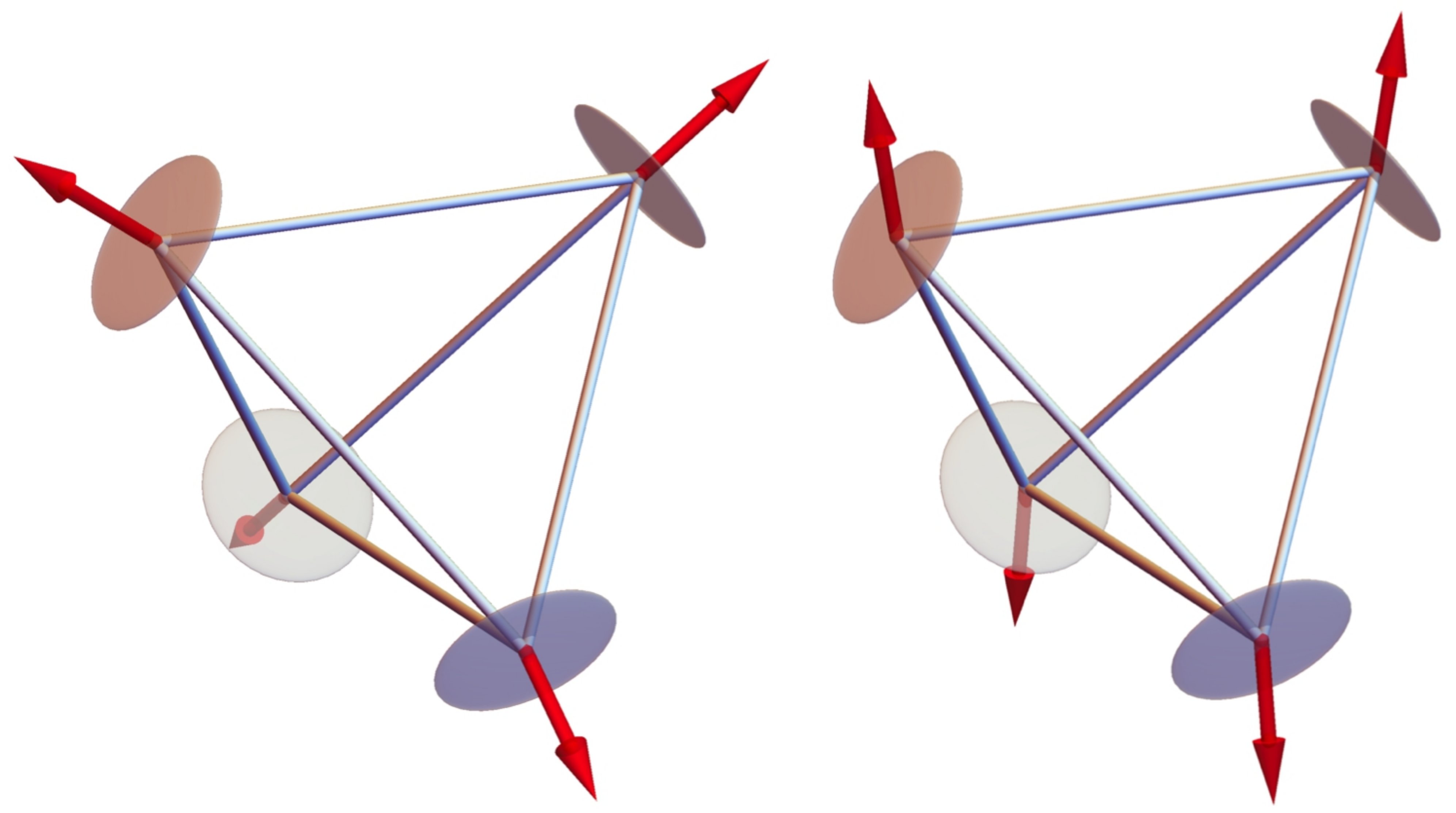}
\end{center}
\caption{left: Pseudo spin picture for the Nd$^{3+}$ ion, showing the measured $z$ component (red) and the transverse $x$ (blue) and $y$ (green) terms. The black arrow corresponds to a moment tilted in the $\tilde z$ direction. right: All out configuration with the spins aligned along $\bf z$ (left) and along $\bf \tilde z$ (right). In the second case, only the projection of the $\tilde z$ component along $\bf z$ is measured. Courtesy of S. Petit.  }
\label{NdZr_pseudospin}    
\end{figure}

The observations in Nd$_2$Zr$_2$O$_7$ can be understood by taking into account the specific nature of the Nd$^{3+}$ ground doublet in the pyrochlore symmetry, which is called a ``dipolar-octupolar" doublet \cite{Huang14}. The Hamiltonian is thus different from the Ising Hamiltonian with a magnetic coupling between the local $z$ components (aligned along $\langle 111 \rangle$ directions) only, used in classical spin ice, and involves transverse terms. In a pseudo spin approach, these transverse terms are associated with $x$ and $y$ components of the pseudo spin, which transforms like a dipole and an octupole respectively, but are not probed by conventional magnetic and neutron measurements, which essentially probe the $z$ component. With this approach, Hamitonian parameters could be refined and reproduce the magnetic excitations. These parameters include a strong coupling between the $x$ components of the pseudo spins, and between the $x$ and $z$ components. The obtained magnetic ground state is a total all in / all out ordering, but with each pseudo spin aligned along a $\bf \tilde z$ direction, tilted from the $\bf z$ direction in the $(\bf x, \bf z)$ plane \cite{Benton16b} (See Figure \ref{NdZr_pseudospin}). In this picture, the measured ordered magnetic moment corresponds to the projection of this $\tilde z$ moment along the $\bf z$ direction, and in the same way, the inelastic scattering corresponds to the projection of the spin waves associated with this $\tilde z$ ordering along the local $\bf z$ direction. This scenario thus precludes fragmentation in the Nd$_2$Zr$_2$O$_7$ ground state. It was however pointed out that fragmentation occurs in the excitations: by linearizing the equations of motion, one can see that the excitations can be separated between a divergence free component, which corresponds to the flat mode shown in Figure \ref{NdZr_INS}, and a propagating charged component, which corresponds to the dispersive modes observed above the flat mode \cite{Benton16b}. 

This model can successfully describe experimental observations in zero field, but cannot describe quantitatively some properties in the presence of a magnetic field, especially the field induced transitions in the ground state \cite{Lhotel15, Lhotel18, xu19a}. 
Interestingly, the pinch point pattern persists above the all in / all out transition \cite{Petit16b}. The nature of this signal - elastic or not - above the transition temperature, together with the temperature dependence of the excitations, is thus of importance to understand the relation between a potential Coulomb phase at high temperature and the all in / all out ordering stabilized at low temperature. In a recent work \cite{Xu20}, the authors propose that the ordering transition is related to the Higgs mechanism where the emergent gauge field of a ``high" temperature Coulomb phase is gapped at the transition by the condensation of emergent monopoles. Further work is needed to confirm this scenario.

\subsubsection{Dimensional reduction in a field}
\label{sec:dimred}
As pointed out in Section \ref{tripod} and Figure \ref{pyrochlore_111}, the pyrochlore lattice can be viewed, along the $[111]$ direction, as a stacking of triangular and kagome layers. In the spin ice case, the spins in the triangular planes are parallel to the $[111]$ axis, and thus almost immediately align when a magnetic field is applied along the $[111]$ directions. The triangular planes being polarized, the ice rule on the tetrahedron is still obeyed provided the magnetic field is not too large (smaller than the effective interactions), so that a degeneracy persists in the kagome plane corresponding to a KI 2 phase on each plane. A two dimensional kagome ice state is thus stabilized. This state is easily identifiable by a magnetization plateau at $1/3$ of the saturated magnetization, and was early observed in Dy$_2$Ti$_2$O$_7$ \cite{Matsuhira02b, Sakakibara03, Moessner03}. This kagome ice state was shown to be robust with respect to a misorientation of the magnetic field \cite{Moessner03, Fennell07, Kao16}. 

This kagome ice state induced by a $[111]$ field in spin ice was thus the first realization of a fragmented state in an experimental system. Diffuse scattering with pinch points, together with the partial ordering of the kagome planes were observed in the canonical spin ices Dy$_2$Ti$_2$O$_7$ and Ho$_2$Ti$_2$O$_7$ long before the fragmentation concept emerged \cite{Tabata06, Fennell07}. The peculiar coexistence of these two features did not attract much attention, maybe due to the additional ordering of the spins in the triangular planes. 

In the peculiar case of the dynamic fragmentation in Nd$_2$Zr$_2$O$_7$, it is amusing that the same dimensional reduction operates for the excitations \cite{Lhotel18}, later confirmed in \cite{xu19a}. The spin ice like flat mode persists when the magnetic field is applied but changes into a kagome ice pattern. This is because the Helmholtz decomposition in the equations of motion can be made within the kagome planes, resulting in the coexistence of a divergence free flat mode with charged quasiparticles propagating in the kagome planes. Interestingly, like in the conventional spin ice case, this dimensional reduction occurs when the triangular spins are polarized ($\mu_0H>0.25$ T), which is well above the field induced transition of the ground state from the all in / all out state to the 3 in - 1 out / 3 out - 1 in state at $\mu_0H\sim0.1$ T.

\section{Openings}
The framework developed here provides a working example of emergent phenomena beyond the traditional picture of spin liquids. Fragmentation offers a platform to investigate emergent gauge fields coupled to a broken symmetry, where the exotic properties of spin liquids take a new flavour. \textcolor{black}{Since an exciting aspect of fragmentation is its experimental observation, it should be considered as a fundamental property of matter. In a nutshell, the take-home message is that even if a system orders, \textit{the} interesting physics might remain hidden in the fluctuations.}

\textcolor{black}{Future directions of research include for example the topological properties of the FSL,} especially unconventional phase transitions beyond the Landau-Ginzburg-Wilson paradigm that would arise when perturbing the FSL. As a starting point, how to include the long-range order in the mapping from the classical problem to the quantum one in $D-1$ dimensions~\cite{Jaubert08,Powell08a,Powell09a} ? Since the divergence-free fluid supports monopole excitations, fragmentation also provides a new tool to manipulate these quasi-particles. The dynamics of monopoles and monopole holes coupled to a divergence-full background is particularly promising. And as more experimental realisations of fragmentation will be discovered, further coupling mechanisms are likely to be exposed, with a magnetic field \cite{Ryzhkin05,Bramwell09,Mostame14a}, or via magneto-electric coupling \cite{khomskii12a}, further neighbour exchange \cite{Rau16b,Udagawa16a}, spin-lattice coupling \cite{Smerald19a} ... Also what happens when the notion of fragmentation is extended to other gauge fields \textcolor{black}{and other quasi-particles}, such as tensor gauge fields \cite{benton16a,prem18a,yan20a}, allowing a connection to fracton physics \cite{Chamon05a,Haah11a,Nandkishore19a} ? On the experimental front, the next stage could be to use hydrostatic and chemical pressure to explore the phase diagram of iridate pyrochlores and Nd-based pyrochlores, and in particular to see if, in the latter, one can bring the divergence-free flat band to lower energies. Modifying the nature of the magnetic ion could also enhance quantum dynamics, both for pyrochlore and kagome materials, an open question for theorists and experimentalists alike. 
\textcolor{black}{The notion of magnetic fragmentation is a recent concept and many open questions remain to be answered, promising an exciting and active future for the field.}



\section*{Conflict of interest}
The authors declare that they have no conflict of interest.

\bibliographystyle{spphys}       
\bibliography{biblio}

\begin{thebibliography}{100}
\providecommand{\url}[1]{{#1}}
\providecommand{\urlprefix}{URL }
\expandafter\ifx\csname urlstyle\endcsname\relax
  \providecommand{\doi}[1]{DOI \discretionary{}{}{}#1}\else
  \providecommand{\doi}{DOI \discretionary{}{}{}\begingroup
  \urlstyle{rm}\Url}\fi

\bibitem{Savary16b}
L.~Savary, L.~Balents, Reports on Progress in Physics \textbf{80}(1), 016502
  (2016).
\newblock \doi{10.1088/0034-4885/80/1/016502}

\bibitem{Knolle18a}
J.~Knolle, R.~Moessner, Annual Review of Condensed Matter Physics
  \textbf{10}(1), 451 (2019).
\newblock \doi{10.1146/annurev-conmatphys-031218-013401}.
\newblock
  \urlprefix\url{https://doi.org/10.1146/annurev-conmatphys-031218-013401}

\bibitem{Anderson72a}
P.W. Anderson, Science \textbf{177}(4047), 393 (1972).
\newblock \doi{10.1126/science.177.4047.393}.
\newblock \urlprefix\url{https://science.sciencemag.org/content/177/4047/393}

\bibitem{NPhys16a}
Nature Physics \textbf{12}(2), 105 (2016).
\newblock \doi{10.1038/nphys3668}.
\newblock \urlprefix\url{http://www.nature.com/articles/nphys3668}

\bibitem{Wen04a}
X.G. Wen, \emph{Quantum Field Theory of Many-Body Systems} (Oxford University
  Press, 2004)

\bibitem{Kitaev06a}
A.~Kitaev, Annals of Physics \textbf{321}(1), 2  (2006).
\newblock \doi{http://dx.doi.org/10.1016/j.aop.2005.10.005}.
\newblock
  \urlprefix\url{http://www.sciencedirect.com/science/article/pii/S0003491605002381}.
\newblock January Special Issue

\bibitem{Castelnovo08}
C.~Castelnovo, R.~Moessner, S.L. Sondhi, Nature \textbf{451}, 42 (2008)

\bibitem{Ryzhkin05}
I.A. Ryzhkin, J. Exp. Theor. Phys. \textbf{101}, 481 (2005)

\bibitem{huse03a}
D.A. Huse, W.~Krauth, R.~Moessner, S.L. Sondhi, Phys. Rev. Lett.
  \textbf{91}(16), 167004 (2003).
\newblock \doi{ARTN 167004}

\bibitem{isakov04b}
S.V. Isakov, K.~Gregor, R.~Moessner, S.L. Sondhi, Phys. Rev. Lett.
  \textbf{93}(16), 167204 (2004).
\newblock \doi{ARTN 167204}

\bibitem{henley05a}
C.L. Henley, Phys. Rev. B \textbf{71}(1), 014424 (2005).
\newblock \doi{ARTN 014424}

\bibitem{Henley10}
C.L. Henley, Annu. Rev. Con. Mat. Phys. \textbf{1}, 179 (2010)

\bibitem{rehn16a}
J.~Rehn, R.~Moessner, Phil. Trans. R. Soc. A \textbf{374}(2075), 20160093
  (2016).
\newblock \doi{10.1098/rsta.2016.0093}.
\newblock
  \urlprefix\url{http://rsta.royalsocietypublishing.org/content/374/2075/20160093}

\bibitem{Harris97}
M.J. Harris, S.T. Bramwell, D.F. McMorrow, T.~Zeiske, K.W. Godfrey, Phys. Rev.
  Lett. \textbf{79}, 2554 (1997)

\bibitem{bramwell01b}
S.T. Bramwell, M.J.P. Gingras, Science \textbf{294}, 1495 (2001)

\bibitem{Ramirez99}
A.P. Ramirez, A.~Hayashi, R.J. Cava, R.~Siddharthan, B.S. Shastry, Nature
  \textbf{399}, 333 (1999)

\bibitem{Brooks14}
M.E. Brooks-Bartlett, S.T. Banks, L.D.C. Jaubert, A.~Harman-Clarke, P.C.W.
  Holdsworth, Phys. Rev. X \textbf{4}, 011007 (2014)

\bibitem{Kosterlitz73a}
J.M. Kosterlitz, D.J. Thouless, Journal of Physics C-Solid State Physics
  \textbf{6}(7), 1181 (1973)

\bibitem{Villain75a}
J.~Villain, Journal de Physique \textbf{36}(6), 581 (1975).
\newblock \doi{10.1051/jphys:01975003606058100}.
\newblock
  \urlprefix\url{http://www.edpsciences.org/10.1051/jphys:01975003606058100}

\bibitem{Faulkner15a}
M.F. Faulkner, S.T. Bramwell, P.C.W. Holdsworth, Phys. Rev. B \textbf{91},
  155412 (2015).
\newblock \doi{10.1103/PhysRevB.91.155412}.
\newblock \urlprefix\url{https://link.aps.org/doi/10.1103/PhysRevB.91.155412}

\bibitem{Borzi13}
R.A. Borzi, D.~Slobinsky, S.A. Grigera, Phys. Rev. Lett. \textbf{111}, 147204
  (2013)

\bibitem{Petit16b}
S.~Petit, E.~Lhotel, B.~Canals, M.C. Hatnean, J.~Ollivier, H.~Mutka,
  E.~Ressouche, A.R. Wildes, M.R. Lees, G.~Balakrishnan, Nature Phys.
  \textbf{12}, 746 (2016)

\bibitem{Lefrancois17}
E.~Lefran{\c c}ois, V.~Cathelin, E.~Lhotel, J.~Robert, P.~Lejay, C.V. Colin,
  B.~Canals, F.~Damay, J.~Ollivier, B.~F\r{a}k, L.~Chapon, R.~Ballou,
  V.~Simonet, Nature Commun. \textbf{8}, 209 (2017)

\bibitem{Paddison16}
J.A.M. Paddison, H.S. Ong, J.O. Hamp, P.~Mukherjee, X.~Bai, M.G. Tucker, N.P.
  Butch, C.~Castelnovo, M.~Mourigal, S.E. Dutton, Nature Commun. \textbf{7},
  13842 (2016)

\bibitem{Canals16}
B.~Canals, I.A. Chioar, V.D. Nguyen, M.~Hehn, D.~Lacour, F.~Montaigne,
  Locatelli, T.O. A., Mente\c{s}, B.~Santos~Burgos, N.~Rougemaille, Nature
  Commun. \textbf{7}, 11446 (2016)

\bibitem{Raban19}
V.~Raban, C.T. Suen, L.~Berthier, P.C.W. Holdsworth, Phys. Rev. B \textbf{99},
  224425 (2019)

\bibitem{denHertog00}
B.C. den Hertog, M.J.P. Gingras, Phys. Rev. Lett. \textbf{84}, 3430 (2000)

\bibitem{Melko04}
R.G. Melko, M.J.P. Gingras, J. Phys.: Condens. Matter \textbf{16}, R1277 (2004)

\bibitem{Pauling35a}
L.~Pauling, Journal of the American Chemical Society \textbf{57}, 2680 (1935)

\bibitem{youngblood81a}
R.W. Youngblood, J.D. Axe, Phys. Rev. B \textbf{23}(1), 232 (1981)

\bibitem{Fennell09}
T.~Fennell, P.P. Deen, A.R. Wildes, K.~Schmalzl, D.~Prabhakaran, A.T.
  Boothroyd, R.J. Aldus, D.F. McMorrow, S.T. Bramwell, Science \textbf{326},
  415 (2009)

\bibitem{isakov05a}
S.V. Isakov, R.~Moessner, S.L. Sondhi, Phys. Rev. Lett. \textbf{95}(21), 217201
  (2005).
\newblock \doi{ARTN 217201}

\bibitem{Jaubert13a}
L.D.C. Jaubert, M.J. Harris, T.~Fennell, R.G. Melko, S.T. Bramwell, P.C.W.
  Holdsworth, Phys. Rev. X \textbf{3}, 011014 (2013).
\newblock \doi{10.1103/PhysRevX.3.011014}.
\newblock \urlprefix\url{http://link.aps.org/doi/10.1103/PhysRevX.3.011014}

\bibitem{Slobinsky19a}
D.~Slobinsky, L.~Pili, R.A. Borzi, Phys. Rev. B \textbf{100}, 020405 (2019).
\newblock \doi{10.1103/PhysRevB.100.020405}.
\newblock \urlprefix\url{https://link.aps.org/doi/10.1103/PhysRevB.100.020405}

\bibitem{Moller09}
G.~M\"{o}ller, R.~Moessner, Phys. Rev. B \textbf{80}, 140409(R) (2009)

\bibitem{Chern11}
G.W. Chern, P.~Mellado, O.~Tchernyshyov, Phys. Rev. Lett. \textbf{106}, 207202
  (2011)

\bibitem{Fulde02a}
P.~Fulde, K.~Penc, N.~Shannon, Annalen der Physik \textbf{11}(12), 892 (2002)

\bibitem{Sala12a}
G.~Sala, C.~Castelnovo, R.~Moessner, S.L. Sondhi, K.~Kitagawa, M.~Takigawa,
  R.~Higashinaka, Y.~Maeno, Phys. Rev. Lett. \textbf{108}, 217203 (2012).
\newblock \doi{10.1103/PhysRevLett.108.217203}.
\newblock
  \urlprefix\url{http://link.aps.org/doi/10.1103/PhysRevLett.108.217203}

\bibitem{McClarty15a}
P.A. McClarty, O.~Sikora, R.~Moessner, K.~Penc, F.~Pollmann, N.~Shannon, Phys.
  Rev. B \textbf{92}, 094418 (2015).
\newblock \doi{10.1103/PhysRevB.92.094418}.
\newblock \urlprefix\url{http://link.aps.org/doi/10.1103/PhysRevB.92.094418}

\bibitem{Henelius16a}
P.~Henelius, T.~Lin, M.~Enjalran, Z.~Hao, J.G. Rau, J.~Altosaar, F.~Flicker,
  T.~Yavors'kii, M.J.P. Gingras, Phys. Rev. B \textbf{93}, 024402 (2016).
\newblock \doi{10.1103/PhysRevB.93.024402}.
\newblock \urlprefix\url{http://link.aps.org/doi/10.1103/PhysRevB.93.024402}

\bibitem{Sen13a}
A.~Sen, R.~Moessner, S.L. Sondhi, Phys. Rev. Lett. \textbf{110}, 107202 (2013).
\newblock \doi{10.1103/PhysRevLett.110.107202}

\bibitem{twengstrom19a}
M.~Twengstrom, P.~Henelius, S.T. Bramwell, Phys. Rev. Research \textbf{2},
  013305 (2020)

\bibitem{Bramwell09}
S.T. Bramwell, S.R. Giblin, S.~Calder, R.~Aldus, D.~Prabhakaran, T.~Fennell,
  Nature \textbf{461}, 956 (2009)

\bibitem{Dunsiger11a}
S.R. Dunsiger, A.A. Aczel, C.~Arguello, H.~Dabkowska, A.~Dabkowski, M.H. Du,
  T.~Goko, B.~Javanparast, T.~Lin, F.L. Ning, H.M.L. Noad, D.J. Singh, T.J.
  Williams, Y.J. Uemura, M.J.P. Gingras, G.M. Luke, Phys. Rev. Lett.
  \textbf{107}, 207207 (2011).
\newblock \doi{10.1103/PhysRevLett.107.207207}.
\newblock
  \urlprefix\url{https://link.aps.org/doi/10.1103/PhysRevLett.107.207207}

\bibitem{chang13a}
L.J. Chang, M.R. Lees, G.~Balakrishnan, Y.J. Kao, A.D. Hillier, Scientific
  Reports \textbf{3}(1), 1 (2013).
\newblock \doi{10.1038/srep01881}.
\newblock \urlprefix\url{https://www.nature.com/articles/srep01881}

\bibitem{nagle66c}
J.F. Nagle, Physical Review \textbf{152}(1), 190 (1966)

\bibitem{Matsuhira02b}
K.~Matsuhira, Z.~Hiroi, T.~Tayama, S.~Takagi, T.~Sakakibara, J. Phys.: Condens.
  Matter \textbf{14}, 559 (2002)

\bibitem{Moessner03}
R.~Moessner, S.L. Sondhi, Phys. Rev. B \textbf{68}, 064411 (2003)

\bibitem{Lhotel18}
E.~Lhotel, S.~Petit, M.C. Hatnean, J.~Ollivier, H.~Mutka, E.~Ressouche,
  M.~Lees, G.~Balakrishnan, Nature Commun. \textbf{9}, 3786 (2018)

\bibitem{Jaubert15b}
L.D.C. Jaubert, SPIN \textbf{5}, 154005 (2015)

\bibitem{benton16a}
O.~Benton, L.D.C. Jaubert, H.~Yan, N.~Shannon, Nature Communications \textbf{7}
  (2016).
\newblock \urlprefix\url{http://dx.doi.org/10.1038/ncomms11572}

\bibitem{prem18a}
A.~Prem, S.~Vijay, Y.Z. Chou, M.~Pretko, R.M. Nandkishore, Physical Review B
  \textbf{98}(16) (2018).
\newblock \doi{10.1103/PhysRevB.98.165140}.
\newblock \urlprefix\url{https://link.aps.org/doi/10.1103/PhysRevB.98.165140}

\bibitem{yan20a}
H.~Yan, O.~Benton, L.D.C. Jaubert, N.~Shannon, Phys. Rev. Lett. \textbf{124},
  127203 (2020).
\newblock \doi{10.1103/PhysRevLett.124.127203}.
\newblock
  \urlprefix\url{https://link.aps.org/doi/10.1103/PhysRevLett.124.127203}

\bibitem{Petrenko14}
O.A. Petrenko, Low Temp. Phys. \textbf{40}, 106 (2014)

\bibitem{Stewart04}
J.R. Stewart, G.~Ehlers, A.S. Wills, S.T. Bramwell, J.S. Gardner, J. Phys.:
  Condens. Matter \textbf{16}, L321 (2004)

\bibitem{Lefrancois19}
E.~Lefran{\c c}ois, L.~Mangin-Thro, E.~Lhotel, J.~Robert, S.~Petit,
  V.~Cathelin, H.E. Fischer, C.V. Colin, F.~Damay, J.~Ollivier, P.~Lejay, L.C.
  Chapon, V.~Simonet, R.~Ballou, Phys. Rev. B \textbf{99}, 060401(R) (2019)

\bibitem{Jaubert15}
L.D.C. Jaubert, O.~Benton, J.G. Rau, J.~Oitmaa, R.R.P. Singh, N.~Shannon, ,
  M.J.P. Gingras, Phys. Rev. Lett. \textbf{115}, 267208 (2015)

\bibitem{Robert15}
J.~Robert, E.~Lhotel, G.~Remenyi, S.~Sahling, I.~Mirebeau, C.~Decorse,
  B.~Canals, S.~Petit, Phys. Rev. B \textbf{92}, 064425 (2015)

\bibitem{Scheie19}
A.~Scheie, J.~Kindervater, S.~Zhang, H.~Changlani, G.~Sala, G.~Ehlers,
  A.~Heinemann, G.S. Tucker, S.~Koohpayeh, C.~Broholm, Multiphase magnetism in
  yb$_2$ti$_2$o$_7$.
\newblock ArXiv:1912.04913

\bibitem{Petit17}
S.~Petit, E.~Lhotel, F.~Damay, P.~Boutrouille, A.~Forget, D.~Colson, Phys. Rev.
  Lett. \textbf{119}, 187202 (2017)

\bibitem{Yan17}
H.~Yan, O.~Benton, L.~Jaubert, N.~Shannon, Phys. Rev. B \textbf{95}, 094422
  (2017)

\bibitem{savary12a}
L.~Savary, L.~Balents, Phys. Rev. Lett. \textbf{108}(3), 037202 (2012).
\newblock \doi{DOI 10.1103/PhysRevLett.108.037202}

\bibitem{savary13a}
L.~Savary, L.~Balents, Phys. Rev. B \textbf{87}, 205130 (2013).
\newblock \doi{10.1103/PhysRevB.87.205130}

\bibitem{Benton16b}
O.~Benton, Phys. Rev. B \textbf{94}, 104430 (2016)

\bibitem{Carrasquilla15}
J.~Carrasquilla, Z.~Hao, R.G. Melko, Nature Commun. \textbf{6}, 7421 (2015)

\bibitem{Huang17}
Y.P. Huang, M.~Hermele, Phys. Rev. B \textbf{95}, 075130 (2017)

\bibitem{Wu19}
K.H. Wu, Y.P. Huang, Y.J. Kao, Phys. Rev. B \textbf{99}, 134440 (2019)

\bibitem{Wang20a}
Y.~Wang, S.~Humeniuk, Y.~Wan, Phys. Rev. B \textbf{101}, 134414 (2020).
\newblock \doi{10.1103/PhysRevB.101.134414}.
\newblock \urlprefix\url{https://link.aps.org/doi/10.1103/PhysRevB.101.134414}

\bibitem{Yan18}
H.~Yan, R.~Pohle, N.~Shannon, Phys. Rev. B \textbf{98}, 140402(R) (2018)

\bibitem{Mizoguchi18}
T.~Mizoguchi, L.D.C. Jaubert, R.~Moessner, M.~Udagawa, Phys. Rev. B
  \textbf{98}, 144446 (2018)

\bibitem{Bergman06a}
D.L. Bergman, G.A. Fiete, L.~Balents, Phys. Rev. B \textbf{73}, 134402 (2006).
\newblock \doi{10.1103/PhysRevB.73.134402}.
\newblock \urlprefix\url{http://link.aps.org/doi/10.1103/PhysRevB.73.134402}

\bibitem{Sikora09a}
O.~Sikora, F.~Pollmann, N.~Shannon, K.~Penc, P.~Fulde, Phys. Rev. Lett.
  \textbf{103}, 247001 (2009).
\newblock \doi{10.1103/PhysRevLett.103.247001}.
\newblock
  \urlprefix\url{http://link.aps.org/doi/10.1103/PhysRevLett.103.247001}

\bibitem{Sikora11a}
O.~Sikora, N.~Shannon, F.~Pollmann, K.~Penc, P.~Fulde, Phys. Rev. B
  \textbf{84}, 115129 (2011).
\newblock \doi{10.1103/PhysRevB.84.115129}.
\newblock \urlprefix\url{http://link.aps.org/doi/10.1103/PhysRevB.84.115129}

\bibitem{Penc04a}
K.~Penc, N.~Shannon, H.~Shiba, Phys. Rev. Lett. \textbf{93}, 197203 (2004).
\newblock \doi{10.1103/PhysRevLett.93.197203}.
\newblock \urlprefix\url{http://link.aps.org/doi/10.1103/PhysRevLett.93.197203}

\bibitem{Ueda05a}
H.~Ueda, H.A. Katori, H.~Mitamura, T.~Goto, H.~Takagi, Phys. Rev. Lett.
  \textbf{94}, 047202 (2005).
\newblock \doi{10.1103/PhysRevLett.94.047202}.
\newblock \urlprefix\url{http://link.aps.org/doi/10.1103/PhysRevLett.94.047202}

\bibitem{Bergman06b}
D.L. Bergman, R.~Shindou, G.A. Fiete, L.~Balents, Phys. Rev. Lett. \textbf{96},
  097207 (2006).
\newblock \doi{10.1103/PhysRevLett.96.097207}.
\newblock \urlprefix\url{http://link.aps.org/doi/10.1103/PhysRevLett.96.097207}

\bibitem{ross11a}
K.A. Ross, L.~Savary, B.D. Gaulin, L.~Balents, Physical Review X \textbf{1},
  021002 (2011)

\bibitem{Kalmeyer87a}
V.~Kalmeyer, R.B. Laughlin, Phys. Rev. Lett. \textbf{59}, 2095 (1987).
\newblock \doi{10.1103/PhysRevLett.59.2095}.
\newblock \urlprefix\url{http://link.aps.org/doi/10.1103/PhysRevLett.59.2095}

\bibitem{Kalmeyer89a}
V.~Kalmeyer, R.B. Laughlin, Phys. Rev. B \textbf{39}, 11879 (1989).
\newblock \doi{10.1103/PhysRevB.39.11879}.
\newblock \urlprefix\url{http://link.aps.org/doi/10.1103/PhysRevB.39.11879}

\bibitem{Grover10a}
T.~Grover, N.~Trivedi, T.~Senthil, P.A. Lee, Phys. Rev. B \textbf{81}, 245121
  (2010).
\newblock \doi{10.1103/PhysRevB.81.245121}.
\newblock \urlprefix\url{https://link.aps.org/doi/10.1103/PhysRevB.81.245121}

\bibitem{Smerald19a}
A.~Smerald, G.~Jackeli, Phys. Rev. Lett. \textbf{122}, 227202 (2019).
\newblock \doi{10.1103/PhysRevLett.122.227202}.
\newblock
  \urlprefix\url{https://link.aps.org/doi/10.1103/PhysRevLett.122.227202}

\bibitem{guruciaga14a}
P.C. Guruciaga, S.A. Grigera, R.A. Borzi, Phys. Rev. B \textbf{90}, 184423
  (2014).
\newblock \doi{10.1103/PhysRevB.90.184423}.
\newblock \urlprefix\url{http://link.aps.org/doi/10.1103/PhysRevB.90.184423}

\bibitem{Melko01}
R.G. Melko, B.C. den Hertog, M.J.P. Gingras, Phys. Rev. Lett. \textbf{87},
  067203 (2001)

\bibitem{Cathelin20}
V.~Cathelin, E.~Lefran\c{c}ois, J.~Robert, P.C. Guruciaga, C.~Paulsen,
  D.~Prabhakaran, P.~Lejay, F.~Damay, J.~Ollivier, B.~F\r{a}k, L.C. Chapon,
  R.~Ballou, V.~Simonet, P.C.W. Holdsworth, E.~Lhotel, Fragmented monopole
  crystal, dimer entropy and coulomb interactions in dy2ir2o7.
\newblock ArXiv:2005.08807

\bibitem{Poole92a}
P.~Poole, F.~Sciortino, U.~Essmann, H.~Stanley, Nature \textbf{360}, 324 (1992)

\bibitem{Katayama03a}
Y.~Katayama, T.~Mizutani, O.~Utsumi, W.and~Shimomura, M.~Yamakata, K.i.
  Funakoshi, Nature \textbf{403}, 170 (2003)

\bibitem{Sastry03a}
S.~Sastry, C.~Austen~Angell, Nature Materials \textbf{2}, 739 (2003)

\bibitem{Brovchenko05a}
I.~Brovchenko, A.~Geiger, A.~Oleinikova, J. Chem. Phys. \textbf{123}, 044515
  (2005)

\bibitem{Lara98}
D.P. na~Lara, J.A. Plascak, Int. J. Mod. Phys. B \textbf{12}, 2045 (1998)

\bibitem{Kaluarachchi17a}
U.S. Kaluarachchi, S.L. Bud'ko, P.C. Canfield, V.~Taufour, Nat. Commun.
  \textbf{8}, 546 (2017)

\bibitem{Kotegawa11a}
H.~Kotegawa, V.~Taufour, D.~Aoki, G.~Knebel, J.~Flouquet, J. Phys. Soc. Jpn.
  \textbf{80}, 083703 (2011)

\bibitem{Castelnovo11a}
C.~Castelnovo, R.~Moessner, S.L. Sondhi, Phys. Rev. B \textbf{84}, 144435
  (2011).
\newblock \doi{10.1103/PhysRevB.84.144435}.
\newblock \urlprefix\url{http://link.aps.org/doi/10.1103/PhysRevB.84.144435}

\bibitem{Giblin11}
S.R. Giblin, S.T. Bramwell, P.C.W. Holdsworth, D.~Prabhakaran, I.~Terry, Nature
  Phys. \textbf{7}, 252 (2011)

\bibitem{zhou11a}
H.D. Zhou, S.T. Bramwell, J.G. Cheng, C.R. Wiebe, G.~Li, L.~Balicas, J.A.
  Bloxsom, H.J. Silverstein, J.S. Zhou, J.B. Goodenough, J.S. Gardner, Nat
  Commun \textbf{2}, 478 (2011)

\bibitem{Poilblanc11a}
D.~Poilblanc, H.~Tsunetsugu, \emph{Introduction to Frustrated Magnetism}
  (Springer (Ed. Lacroix, Mendels \& Mila), 2011), \emph{Solid State Science},
  vol. 164, chap. Mobile Holes in Frustrated Quantum Magnets and Itinerant
  Fermions on Frustrated Geometries

\bibitem{skjaervo19a}
S.H. Skj{\ae}rv{\o}, C.H. Marrows, R.L. Stamps, L.J. Heyderman, Nature Reviews
  Physics pp. 1--16 (2019).
\newblock \doi{10.1038/s42254-019-0118-3}.
\newblock \urlprefix\url{http://www.nature.com/articles/s42254-019-0118-3}

\bibitem{Rougemaille19}
N.~Rougemaille, B.~Canals, Eur. Phys. J. B. \textbf{92}(62) (2019)

\bibitem{Zhang13}
S.~Zhang, I.~Gilbert, C.~Nisoli, G.W. Chern, M.J. Erickson, L.~O'Brien,
  C.~Leighton, P.E. Lammert, V.H. Crespi, P.~Schiffer, Nature \textbf{500}, 553
  (2013)

\bibitem{Dun16}
Z.L. Dun, J.~Trinh, K.~Li, M.~Lee, K.W. Chen, R.~Baumbach, Y.F. Hu, Y.X. Wang,
  E.S. Choi, B.S. Shastry, A.P. Ramirez, H.D. Zhou, Phys. Rev. Lett.
  \textbf{116}, 157201 (2016)

\bibitem{Dun17}
Z.L. Dun, J.~Trinh, M.~Lee, E.S. Choi, K.~Li, Y.F. Hu, Y.X. Wang, N.~Blanc,
  A.P. Ramirez, H.D. Zhou, Phys. Rev. B \textbf{95}, 104439 (2017)

\bibitem{Dun18}
Z.~Dun, X.~Bai, J.A.M. Paddison, N.P. Butch, C.D. Cruz, M.B. Stone, T.~Hong,
  M.~Mourigal, H.~Zhou, Quantum spin fragmentation in kagome ice
  ho$_3$mg$_2$sb$_3$o$_{14}$.
\newblock ArXiv1806.04081

\bibitem{Wang68}
Y.L. Wang, B.R. Cooper, Phys. Rev. \textbf{172}, 539 (1968)

\bibitem{Dun19a}
Z.~Dun, X.~Bai, J.A.M. Paddison, E.~Hollingworth, N.P. Butch, C.D. Cruz, M.B.
  Stone, T.~Hong, F.~Demmel, M.~Mourigal, H.~Zhou.
\newblock Quantum versus classical spin fragmentation in dipolar kagome ice
  ho3mg2sb3o14 (2019)

\bibitem{Scheie16}
A.~Scheie, M.~Sanders, J.~Krizan, Y.~Qiu, R.J. Cava, C.~Broholm, Phys. Rev. B
  \textbf{93}, 180407(R) (2016)

\bibitem{Matsuhira11}
K.~Matsuhira, C.~Paulsen, E.~Lhotel, C.~Sekine, Z.~Hiroi, S.~Takagi, J. Phys.
  Soc. Jap. \textbf{80}, 123711 (2011)

\bibitem{Tomiyasu12}
K.~Tomiyasu, K.~Matsuhira, K.~Iwasa, M.~Watahiki, S.~Takagi, M.~Wakeshima,
  Y.~Hinatsu, M.~Yokoyama, K.~Ohoyama, K.~Yamada, J. Phys. Soc. Jap.
  \textbf{81}, 034709 (2012)

\bibitem{Lefrancois15}
E.~Lefran{\c c}ois, V.~Simonet, R.~Ballou, E.~Lhotel, A.~Hadj-Azzem,
  S.~Kodjikian, P.~Lejay, P.~Manuel, D.~Khalyavin, L.C. Chapon, Phys. Rev.
  Lett. \textbf{114}, 247202 (2015)

\bibitem{Gardner10}
J.S. Gardner, M.J.P. Gingras, J.E. Greedan, Rev. Mod. Phys. \textbf{82}(53)
  (2010)

\bibitem{Castelnovo12}
C.~Castelnovo, R.~Moessner, S.~Sondhi, Annu. Rev. Condens. Matter Phys.
  \textbf{3}, 35 (2012)

\bibitem{Ciomaga15}
M.C. Hatnean, M.R. Lees, O.A. Petrenko, D.S. Keeble, G.~Balakrishnan, M.J.
  Gutmann, V.V. Klekovkina, B.Z. Malkin, Phys. Rev. B \textbf{91}, 174416
  (2015)

\bibitem{Lhotel15}
E.~Lhotel, S.~Petit, S.~Guitteny, O.~Florea, M.~Ciomaga~Hatnean, C.~Colin,
  E.~Ressouche, M.R. Lees, G.~Balakrishnan, Phys. Rev. Lett. \textbf{115},
  197202 (2015)

\bibitem{Xu15}
J.~Xu, V.K. Anand, A.K. Bera, M.~Frontzek, D.L. Abernathy, N.~Casati,
  K.~Siemensmeyer, B.~Lake, Phys. Rev. B \textbf{92}, 224430 (2015)

\bibitem{Anand15}
V.K. Anand, A.K. Bera, J.~Xu, T.~Herrmannsd\"orfer, C.~Ritter, B.~Lake, Phys.
  Rev. B \textbf{92}, 184418 (2015)

\bibitem{Bertin15}
A.~Bertin, P.D. de~R\'eotier, B.~F\r{a}k, C.~Marin, A.~Yaouanc, A.~Forget,
  D.~Sheptyakov, B.~Frick, C.~Ritter, A.~Amato, C.~Baines, P.J.C. King, Phys.
  Rev. B \textbf{92}, 144423 (2015)

\bibitem{Huang14}
Y.P. Huang, G.~Chen, M.~Hermele, Phys. Rev. Lett. \textbf{112}, 167203 (2014)

\bibitem{xu19a}
J.~Xu, O.~Benton, V.K. Anand, A.T.M.N. Islam, T.~Guidi, G.~Ehlers, E.~Feng,
  Y.~Su, A.~Sakai, P.~Gegenwart, B.~Lake, Phys. Rev. B \textbf{99}, 144420
  (2019).
\newblock \doi{10.1103/PhysRevB.99.144420}.
\newblock \urlprefix\url{https://link.aps.org/doi/10.1103/PhysRevB.99.144420}

\bibitem{Xu20}
J.~Xu, O.~Benton, A.T.M.N. Islam, T.~Guidi, G.~Ehlers, B.~Lake, Phys. Rev.
  Lett. \textbf{124}, 097203 (2020)

\bibitem{Sakakibara03}
T.~Sakakibara, T.~Tayama, Z.~Hiroi, K.~Matsuhira, S.~Takagi, Phys. Rev. Lett.
  \textbf{90}, 207205 (2003)

\bibitem{Fennell07}
T.~Fennell, S.T. Bramwell, D.F. McMorrow, P.~Manuel, A.R. Wildes, Nature Phys.
  \textbf{3}, 566 (2007)

\bibitem{Kao16}
W.H. Kao, P.C.W. Holdsworth, Y.J. Kao, Phys. Rev. B \textbf{93}, 180410(R)
  (2016)

\bibitem{Tabata06}
Y.~Tabata, H.~Kadowaki, K.~Matsuhira, Z.~Hiroi, N.~Aso, E.~Ressouche,
  B.~F\r{a}k, Phys. Rev. Lett. \textbf{97}, 257205 (2006)

\bibitem{Jaubert08}
L.D.C. Jaubert, J.T. Chalker, P.C.W. Holdsworth, R.~Moessner, Phys. Rev. Lett.
  \textbf{100}, 067207 (2008)

\bibitem{Powell08a}
S.~Powell, J.T. Chalker, Phys. Rev. B \textbf{78}(2), 024422 (2008).
\newblock \doi{ARTN 024422}

\bibitem{Powell09a}
S.~Powell, J.T. Chalker, Phys. Rev. B \textbf{80}, 134413 (2009).
\newblock \doi{10.1103/PhysRevB.80.134413}.
\newblock \urlprefix\url{http://link.aps.org/doi/10.1103/PhysRevB.80.134413}

\bibitem{Mostame14a}
S.~Mostame, C.~Castelnovo, R.~Moessner, S.L. Sondhi, Proceedings of the
  National Academy of Sciences \textbf{111}(2), 640 (2014).
\newblock \doi{10.1073/pnas.1317631111}.
\newblock \urlprefix\url{http://www.pnas.org/content/111/2/640.abstract}

\bibitem{khomskii12a}
D.~Khomskii, Nature Communications \textbf{3}, 904 (2012)

\bibitem{Rau16b}
J.G. Rau, M.J.P. Gingras, Nature Communications \textbf{7}, ncomms12234 (2016).
\newblock \doi{10.1038/ncomms12234}.
\newblock \urlprefix\url{https://www.nature.com/articles/ncomms12234}

\bibitem{Udagawa16a}
M.~Udagawa, L.D.C. Jaubert, C.~Castelnovo, R.~Moessner, Physical Review B
  \textbf{94}(10), 104416 (2016).
\newblock \doi{10.1103/PhysRevB.94.104416}.
\newblock \urlprefix\url{https://link.aps.org/doi/10.1103/PhysRevB.94.104416}

\bibitem{Chamon05a}
C.~Chamon, Phys. Rev. Lett. \textbf{94}, 040402 (2005).
\newblock \doi{10.1103/PhysRevLett.94.040402}.
\newblock
  \urlprefix\url{https://link.aps.org/doi/10.1103/PhysRevLett.94.040402}

\bibitem{Haah11a}
J.~Haah, Phys. Rev. A \textbf{83}, 042330 (2011).
\newblock \doi{10.1103/PhysRevA.83.042330}.
\newblock \urlprefix\url{https://link.aps.org/doi/10.1103/PhysRevA.83.042330}

\bibitem{Nandkishore19a}
R.M. Nandkishore, M.~Hermele, Annual Review of Condensed Matter Physics
  \textbf{10}(1), 295 (2019).
\newblock \doi{10.1146/annurev-conmatphys-031218-013604}.
\newblock
  \urlprefix\url{https://doi.org/10.1146/annurev-conmatphys-031218-013604}

\end{thebibliography}

\end{document}